\documentclass[prd,aps,showpacs,tightenlines,10pt, reprint,nofootinbib,superscriptaddress,floatfix,twocolumn]{revtex4-2}

\usepackage[utf8]{inputenc} 

\usepackage{acronym}
\usepackage{amsmath}
\usepackage{amsthm}  
\usepackage{amssymb}
\usepackage{mathtools}
\usepackage{phaistos}
\usepackage{graphicx}
\usepackage[section]{placeins}

\usepackage[dvipsnames]{xcolor}   

\usepackage{physics} 
\usepackage{IEEEtrantools}  
\usepackage{xspace}
\usepackage{cancel}
\usepackage[normalem]{ulem} 

\usepackage{tikz}
\usepackage[compat=1.1.0]{tikz-feynman}

\tikzset{graviton/.style={decorate, decoration={snake, amplitude=.4mm, segment length=1.5mm, pre length=.5mm, post length=.5mm}, double}}
\usepackage{array}
\newcolumntype{C}[1]{>{\centering\let\newline\\\arraybackslash\hspace{0pt}}m{#1}}

\usepackage{cellspace}
\usepackage{comment}
\usepackage{epsfig}
\usepackage{multirow} 
\usepackage{color}
\usepackage[export]{adjustbox} 
\usepackage{floatrow}
\usepackage{float}
\usepackage[font=small,labelfont=bf,justification=Justified]{caption}
\usepackage{subcaption}
\usepackage{makecell}
\usepackage[titles]{tocloft}
\usepackage{titlesec}

\titleformat{\section}
  {\centering\normalfont\fontsize{10}{15}\bfseries}{\thesection}{1em}{}
\titleformat{\subsection}
  {\centering\normalfont\fontsize{10}{15}\bfseries}{\thesubsection}{1em}{}

\usepackage{verbatim}
\newfloatcommand{capbtabbox}{table}[][\FBwidth]
\usepackage{tabularx}
\usepackage[pdfencoding=auto, psdextra]{hyperref}

\usepackage{units}
\usepackage[T1]{fontenc}
\usepackage[normalem]{ulem} 
\usepackage{slashed}
\usepackage{bbold}
\usepackage{orcidlink}

\usepackage{academicons}
\definecolor{orcidlogocol}{HTML}{A6CE39}
\newcommand{\orcid}[1]{\href{https://orcid.org/#1}{\textcolor[HTML]{A6CE39}{\aiOrcid}}}

\hypersetup{
    colorlinks,
    linkcolor={red!60!black},
    citecolor={blue!50!black},
    urlcolor={blue!80!black}
}
\allowdisplaybreaks

\tikzset{
    scalarmoredash/.style={decorate, draw=black, dashed, dash pattern=on 0.5pt off 0.5pt}
}
\tikzset{
    scalarlessdash/.style={decorate, draw=black, dashed, dash pattern=on 6pt off 2pt}
 }
\tikzset{
    potentialScalar/.style={decorate, draw=black, dashed, dash pattern=on 4pt off 2pt, double}
}
\tikzset{
    radiationScalar/.style={decorate, draw=black, dashed, dash pattern=on 4pt off 2pt}
}
\tikzset{
    radiationGraviton/.style={photon}
}
\tikzset{
    potentialGraviton/.style={decorate, decoration={snake, amplitude=.4mm, segment length=2mm, pre length=.5mm, post length=.5mm}, double}
}

\advance\cftsecnumwidth 0.5em\relax
\advance\cftsubsecindent 0.5em\relax
\advance\cftsubsecnumwidth 0.5em\relax

\newcommand\cincludegraphics[2][]{\raisebox{-0.4\height}{\includegraphics[#1]{#2}}}

\newcommand{\PhiBar}{\Phi}
\newcommand{\phiBar}{\Bar{\phi}}

\newcommand{\SEH}{\mathcal{S}_\mathrm{EH}}
\newcommand{\SGF}{\mathcal{S}_\mathrm{GF}}
\newcommand{\SPhi}{\mathcal{S}_\Phi}
\newcommand{\Spp}{\mathcal{S}_\mathrm{pp}}
\newcommand{\Seff}{\mathcal{S}_\mathrm{eff}}

\newcommand{\MPl}{M_\mathrm{Pl}}
\newcommand{\vbf}{\boldsymbol{v}}
\newcommand{\kbf}{{\boldsymbol{k}}}
\newcommand{\kabsbf}{|\boldsymbol{k}|}
\newcommand{\qabsbf}{|\boldsymbol{q}|}

\newcommand{\xbf}{\boldsymbol{x}}
\newcommand{\qbf}{\boldsymbol{q}}
\newcommand{\rbf}{\boldsymbol{r}}
\newcommand{\nbf}{\boldsymbol{n}}

\newcommand{\sumAB}{\sum_{1 \leftrightarrow 2}}
\newcommand{\sumi}{\sum_{i = 1,2}}

\newcommand{\LEIH}{L_\mathrm{EIH}}
\newcommand{\Ei}{{\rm Ei}}

\newcommand{\hmunu}{h_{\mu\nu}}
\newcommand{\gmunu}{g_{\mu\nu}}
\newcommand{\hbarmunu}{\bar{h}_{\mu\nu}}
\newcommand{\Hmunu}{H_{\mu\nu}}
\newcommand{\Hkbfmunu}{H_{\kbf\mu\nu}}

\newcommand{\Qbar}{\Bar{Q}}
\newcommand{\msbar}{\Bar{m}_s}

\begin{document}

\title{Binary systems in massive scalar-tensor theories: Next-to-leading order gravitational wave phase from effective field theory}

\author{Robin Fynn Diedrichs\,\orcidlink{0009-0007-3998-4608}}
\email{diedrichs@itp.uni-frankfurt.de} 
\affiliation{Institute for Theoretical Physics, Goethe University, 60438 Frankfurt am Main, Germany}

\author{Daniel Schmitt\,\orcidlink{0000-0003-3369-2253}}
\email{dschmitt@itp.uni-frankfurt.de} 
\affiliation{Institute for Theoretical Physics, Goethe University, 60438 Frankfurt am Main, Germany}

\author{Laura Sagunski\,\orcidlink{0000-0002-3506-3306}}
\email{sagunski@itp.uni-frankfurt.de} 
\affiliation{Institute for Theoretical Physics, Goethe University, 60438 Frankfurt am Main, Germany}

\date{\today}
\begin{abstract}
Neutron star binaries and their associated gravitational wave signal facilitate precision tests of General Relativity. 
Any deviation of the detected gravitational waveform from General Relativity would therefore be a smoking gun signature of new physics, in the form of additional forces, dark matter particles, or extra gravitational degrees of freedom. To be able to probe new theories, precise knowledge of the expected waveform is required. In our work, we consider a generic setup by augmenting General Relativity with an additional, massive scalar field. We then compute the inspiral dynamics of a binary system, for circular orbits, by employing an effective field theoretical approach, while giving a detailed introduction to the computational framework. Finally, we derive the modified TaylorF2 phase of the gravitational wave signal at next-to-leading order in the post-Newtonian expansion, and leading order in the parameters of the scalar sector, such as the scalar charge. As a consequence of our model-agnostic approach, our results are readily adaptable to a plethora of new physics scenarios, including modified gravity theories and scalar dark matter models. 
\end{abstract}

\maketitle

\tableofcontents        

\section{Introduction \label{sec:intro}} 
The first direct detection of gravitational waves (GWs) from a neutron star~(NS) binary \cite{LIGOScientific:2017vwq}---GW170817---has opened new pathways in modern astrophysics by providing further precision tests of General Relativity~(GR)~\cite{LIGOScientific:2018dkp}. 
As possible new physics effects on the neutron star properties\footnote{Black hole~(BH)~binaries, in contrast, are typically not as sensitive to new physics as a consequence of no-hair theorems; see e.g.~Ref.~\cite{Herdeiro:2015waa} for a review.} would inherently affect the emission of gravitational radiation, deviations from the GR expectation would distinctly point toward new phenomena. This renders NS binary systems a promising cosmic laboratory. 

Besides existing GW data from the LIGO/Virgo collaboration, upcoming data from future runs and next-generation observatories---such as the Laser Interferometer Space Antenna~(LISA)~\cite{2017arXiv170200786A} or the Einstein Telescope~(ET)~\cite{Punturo:2010zz}---will be available to significantly narrow down the landscape of new theories. This concerns both modifications of GR~\cite{Capozziello:2011et, Quartin:2023tpl, Higashino:2022izi} and extensions of the Standard Model (SM) of particle physics \cite{ParticleDataGroup:2022pth}. Such models are in general well motivated, as they can, e.g., reproduce the phenomenology of dark matter \cite{Bertone:2016nfn} or cosmic inflation~\cite{Nojiri:2017ncd,Lyth:1998xn}.

In order to constrain new physics with GW observations, however, it is crucial to provide precise theoretical predictions of the gravitational waveform. To this end, different methods may be employed. While the late stages of the binary evolution are only accessible via non-perturbative techniques such as numerical relativity, the early inspiral can be tackled analytically. This is typically done with a post-Newtonian (PN) expansion~\cite{Blanchet:2013haa}, valid at small velocities.

In this work, we use an effective field theory~(EFT)~approach to model the inspiral dynamics \cite{Goldberger:2004jt,Goldberger:2007hy} of two compact objects. Here, the scale hierarchy between the constituent size, the orbital separation, and the wavelength of the emitted GWs is employed to systematically compute the waveform at the desired order in the PN expansion. This framework has been applied extensively within pure GR~\cite{Galley:2009px,Porto:2016pyg,Foffa:2019yfl,Kalin:2020fhe,Dlapa:2021npj,Goldberger:2009qd,Goldberger:2012kf,Goldberger:2022ebt,Goldberger:2022rqf}, successfully reproducing results obtained by more traditional methods.
More recently, this has been extended to binaries in certain modified gravities \cite{Cardoso:2008gn,Kuntz:2019zef, Poddar:2021yjd} or in the presence of dark matter \cite{Huang:2018pbu,Zhang:2021mks,Bhattacharyya:2023kbh}.

Here, we take a step further and provide a generalized computation of the binary evolution in massive scalar-tensor theories,~i.e., GR augmented by a massive scalar field. 
By choosing the most general scalar potential and interactions, we construct the EFT such that all results may be adapted to a variety of theories exhibiting additional scalar degrees of freedom.
Hence, our calculation extends previous works on binary dynamics in scalar-tensor gravity \cite{Bernard:2018hta,Bernard:2018ivi,Bernard:2019yfz,Bernard:2022noq,Julie:2019sab,Shiralilou:2020gah,Shiralilou:2021mfl,Julie:2022qux,Bernard:2023eul, Mirshekari:2013vb, Lang:2013fna, Lang:2014osa, Sennett:2016klh,Alsing:2011er,Berti:2012bp}.
We compute the conservative and dissipative dynamics of the system. 
Moreover, we derive the power loss via scalar radiation.
Finally, we arrive at our final result---the phase of the gravitational waveform at next-to-leading (NLO) order. 
This comprises, e.g., binary systems in $f(R)$ models \cite{Sagunski:2017nzb} or scalar dark matter clouds~\cite{Diedrichs:2023trk}, to name just a few. All our calculations are documented in a detailed manner, allowing one to easily reproduce our results and extend the computation to higher precision. In addition, we provide a Python implementation of the final waveform phase in the TaylorF2 approximation~\cite{waveform_code_snippet}.

In Sec.~\ref{sec:scalarTensorTheory}, we introduce our setup. 
Subsequently, the most important aspects of the EFT approach are reviewed in Sec.~\ref{sec:EFT}. Then, we construct the EFT for a generic scalar-tensor theory in Sec.~\ref{sec:scalar_tensor_eft}, including the computation of the power loss. Section~\ref{sec:waveform} contains the resulting TaylorF2 phase. 
Technical aspects, as well as computational details, are discussed in the appendixes~\ref{app:redundant_operators}~--~\ref{app:feynman_diagrams}.

In the following, we use units in which $\hbar=c=1$ unless stated otherwise. We further employ $\MPl^2 = 1/(32\pi G)$ and $\eta_{\mu\nu} = \text{diag}\, (+1, -1, -1, -1)$.


\section{Scalar-Tensor Theory} \label{sec:scalarTensorTheory}
In this work, we consider scalar extensions of General Relativity. 
In the Einstein frame and in the presence of a binary system, such theories are described by an action of the form \cite{Capozziello:2011et}
\begin{equation} \label{eq:action}    
\mathcal{S} = \SEH + \SGF + \Spp + \SPhi \; .
\end{equation}
Here,
\begin{equation} \label{eq:SEH}
    \SEH = -2 \MPl^2 \int d^4x \sqrt{-g} R 
\end{equation}
is the Einstein-Hilbert action with the Ricci scalar $R$ and the metric determinant $g = \det \gmunu$. $\SGF$ is a gauge-fixing term that we may choose at our convenience to simplify calculations. To this end, we employ the harmonic gauge \cite{Goldberger:2004jt}
\begin{equation}
    \SGF = \MPl^2 \int d^4x \sqrt{-g} \,\Gamma^\mu \Gamma^\nu g_{\mu\nu} \; .
\end{equation}
Note that we do not include ghost fields as we conduct our computations at the classical level.

The pure GR action is augmented by a real, massive, and self-interacting scalar field:
\begin{equation} \label{eq:scalar_action}
\SPhi = \int d^4x \sqrt{-g} \left[ \frac{1}{2} \partial_\mu \phi \partial^\mu \phi - V(\phi) \right] \; .
\end{equation}
Here the first term denotes the kinetic term, while $V(\phi)$ is the scalar potential which can take different functional forms, depending on the model. The most general ansatz, expanded around the potential minimum, then reads
\begin{equation} \label{eq:scalar_potential}
    V(\phi) = \frac{m_s^2}{2!} \phi^2 + \MPl \frac{c_3}{3!} \phi^3 + \frac{\lambda}{4!} \phi^4 + ... \; ,
\end{equation}
where $m_s$ denotes the scalar mass, and $c_3$ parametrizes the cubic self-interaction. The quartic coupling $\lambda$, as well as higher-order self-interactions, have no effect at the PN order we consider in the following, therefore will be dropped henceforth. 
Note that we may take the massless limit, $m_s \rightarrow 0$, such that the results obtained in this work also apply to massless scalar fields.\footnote{The cubic interaction term typically leads to IR divergences in the absence of a mass term when evaluating tree-level interactions, as we are doing here. These divergences are of the form $\propto c_3 / m_s$ and do thus not have to pose a problem, since in many theories $c_3$ is proportional to powers of the mass. See App.~\ref{app:validity} for an example. 
Further note that in the case of a quantum theory with no bare mass term, one generally expects a generation of a scalar mass term $m_s \sim c_3$ from loop diagrams. See also Ref.~\cite{Porto:2007pw} for a discussion on this.}

This new scalar degree of freedom can, e.g.,~arise from modifications of GR \cite{Damour:1998jk,Palenzuela:2013hsa,Niu:2021nic,Creci:2023cfx,Ma:2023sok} such as $f(R)$~gravity~\cite{Sotiriou:2008rp,DeFelice:2010aj,Sagunski:2017nzb}, or SM extensions~\cite{Hook:2017psm,Sagunski:2017nzb,Zhang:2021mks,Bhattacharyya:2023kbh}. To retain model independence, we leave the coefficients $m_s,~c_3,$ and $\lambda$ generic. Then, our results can directly be applied to specific theories by adapting the scalar potential. A discussion of the validity of our approach is given in Appendix~\ref{app:validity}.

As long as the distance between the compact objects is much larger than their size, the binary constituents are---to leading order (LO)---well described by point particles. The corresponding point-particle action reads
\begin{equation} \label{eq:Spp}
    \Spp = - \sum_{n = 1,2} \int d\tau \left(M_n + q_n \frac{\phi}{\MPl} + p_n \left(\frac{\phi}{\MPl}\right)^2 + ...\right) \; ,
\end{equation}
where the line element is
\begin{equation} \label{eq:line_element}
    d\tau = \left(g_{\mu\nu} dx^\mu dx^\nu \right)^{1/2} \; .
\end{equation}
Further, we decompose the metric via
\begin{equation}
    g_{\mu \nu} = \eta_{\mu\nu} +\frac{h_{\mu\nu}}{\MPl} \; ,
\end{equation}
where $\hmunu$ is a perturbation on top of the flat Minkowski metric $\eta_{\mu\nu}$.

The first term in Eq.~\eqref{eq:Spp} simply corresponds to the action of a point particle in pure GR.
The $\phi$-dependent terms denote the scalar contributions, which parametrize the possible couplings of the scalar field to the worldlines of the NSs. The coupling coefficients are denoted as the \textit{scalar charge} $q_n$ and \textit{induced scalar charge} $p_n$. Similar to our treatment of the scalar potential, we include these EFT parameters in a model-agnostic manner. Our results can be mapped to specific models by adjusting Eq.~\eqref{eq:scalar_potential} and matching $q_n$ and $p_n$ to the respective full theory.

Note that we left out some operators in Eq.~\eqref{eq:Spp}, e.g.~$u^\mu \partial_\mu \phi$, with $u^\mu = dx^\mu / d\tau$ being the four-velocity. This term is equal to a total derivative and is thus not relevant. 
A list of redundant operators is given in App.~\ref{app:redundant_operators}, where we explain that, at the order we consider in this work, we do not have to consider any other operators than those already listed in the point-particle action above.

Please note that Eq.~\eqref{eq:Spp} generally also contains finite-size contributions from pure GR at higher orders in the expansion. These, however, only contribute from the fifth PN order onward for non-spinning objects, thus are not relevant here and will be neglected in the further discussion\footnote{The first two operators that enter are $E^{\mu\nu}E_{\mu\nu}$ and $B^{\mu\nu}B_{\mu\nu}$ with $E_{\mu\nu} = R_{\mu\alpha\nu\beta} u^\alpha u^\beta$ and $B_{\mu\nu} = \epsilon_{\mu\alpha\beta\rho} R^{\alpha\beta}{}_{\sigma\nu} u^\sigma u^\rho$, which are the decompositions of the Riemann tensor into its electric and magnetic type components, respectively. The Wilson coefficients of these operators capture the leading order finite size effects arising in pure GR, i.e.,~the tidal deformability.}.

It is useful to compare this setup with the one typically used in Brans-Dicke (BD) theory. There, the point-particle action takes the shape
\begin{equation}
    \mathcal{S}_{pp} = - \sum_{n = 1,2} \int d \tau M_n(\varphi),
\end{equation}
with $\varphi$ being the BD field and $M_n(\varphi)$ being the field-dependent mass. One then further defines the so-called \textit{sensitivities}, with the first two given as
\begin{align}
    s_n = \frac{d \ln M_n}{d \ln \varphi} \,,\quad \text{and} \quad s'_n = \frac{d^2 \ln M_n}{(d \ln \varphi)^2}\,.
\end{align}
The sensitivities are related to the scalar (induced) charges via
\begin{align}
\label{BDtoSTdictionary}
    q_n &= \frac{M_{n,0}}{\sqrt{6 + 4 \omega_0}} \left( s_n - \frac{1}{2} \right), \\
    p_n &= \frac{M_{n,0}}{12 + 8 \omega_0} \left[s'_n +  \left(s_n - \frac{1}{2}\right)^2 \right],
\end{align}
with $M_{n,0} = M_n(\varphi_0)$, $\omega_0 = \omega(\varphi_0)$, $\varphi_0$ denoting the asymptotic value of the BD field and $\omega$ denoting the BD coupling constant. These relations allow for a comparison of our results -- in the limit of $c_3 = 0$ and $m_s=0$ -- with previous computations for massless BD theory \cite{Bernard:2022noq, Sennett:2016klh}. To this end, a field transformation of the form $\phi = \log\left(\varphi/\varphi_0\right)$ is required. 
Therefore, an expansion of around $\phi = 0$ is equivalent to an expansion of $\varphi$ around $\varphi_0$.

\section{Effective Field Theory Approach to Binary Systems} \label{sec:EFT}
Let us now outline the most important features of the EFT method to study binary systems in GR and beyond. This discussion comprises the two main ingredients to construct the effective theory: the \textit{separation of scales} during the inspiral and the resulting \textit{power counting scheme}. We closely follow Refs.~\cite{Goldberger:2004jt,Goldberger:2007hy,Porto:2016pyg,Goldberger:2022ebt,Goldberger:2022rqf}.\\

\paragraph*{\bf Separation of scales.} The first step in the EFT construction is to identify the different energy scales during the inspiral. We consider a NS binary system with constituents of typical size $R \sim R_s$, where $R_s$ is the Schwarzschild radius, separated by a distance~$r$. The wavelength of the gravitational radiation detected on Earth is $\lambda$. 
Denoting the typical velocity of the binary system by $v$, we obtain using the virial theorem
\begin{equation} \label{eq:velocity_estimate}
    v^2 \sim \frac{G M}{r} \equiv \frac{R_s}{2 r} \; .
\end{equation}
This holds as long as the dynamics is governed to LO by a Newtonian interaction. Here we have introduced the Schwarzschild radius $R_s = 2GM$, where $M$ is the mass of the constituents and $G$ is Newton's constant.

From Eq.~\eqref{eq:velocity_estimate} we see that while the distance between the constituents is much larger than their size $r \gg R_s$, the velocity of the system is small $v \ll 1$. This corresponds to the early stages of the binary evolution. The typical wavelength of the emitted GWs relates to the orbital frequency via $f \sim v/r$. Thus, there exists a clear separation of scales during the inspiral,
\begin{equation}
\lambda \sim r/v \gg r \gg R_s \; .     
\end{equation}
This hierarchy will be employed to integrate out the \textit{hard scale} $r$ in order to construct a theory valid at the \textit{soft scale} $\lambda$. 

Let us consider the typical momenta present at the hard and soft scale. The long-wavelength modes carry away energy from the binary system in the form of gravitational and scalar radiation.\footnote{Note that scalar radiation is only emitted if the scalar mass is sufficiently small, i.e., $m_s < \omega$ where $\omega$ is the orbital frequency. This will become apparent when computing the radiative dynamics in Sec.~\ref{subsec::radiative_dynamics}.} To appear as propagating degrees of freedom, the fields must be on shell and therefore the components of its four-momentum scale as
\begin{equation}
    k^0 \sim \frac{v}{r} \; , \qquad |\kbf| \sim \frac{v}{r}  \; . 
\end{equation}
In addition to the radiation modes, the system contains short-wavelength degrees of freedom contributing to the binding energy between the constituents. 
These hard modes naturally live on the orbital separation scale $\sim r$ and therefore carry momenta
\begin{equation}
    k^0 \sim \frac{v}{r} \; , \qquad |\kbf| \sim \frac{1}{r} \; .
\end{equation}
Those off-shell modes do not appear as propagating degrees of freedom.

To make the distinction between the short- and long-wavelength scales explicit, we decompose the graviton and scalar fields as
\begin{equation}
    \begin{split}
    \hmunu(x) &= \hbarmunu (x)+ \Hmunu(x) \; , \\
    \phi(x) &= \Bar{\phi}(x) + \Phi (x)  \; .
    \end{split}
\end{equation}
Here, $\Hmunu(x)$ and $\Phi(x)$ denote the hard modes which mediate the interaction between the binaries at the scale $r$. Conversely, $\hbarmunu (x)$ and $\Bar{\phi}(x)$ represent the soft, propagating degrees of freedom.
From the scaling relations of the momenta, we find the scaling of derivatives acting on the respective modes 
\begin{equation}
    \begin{split}
       \partial_0 \Hmunu &= \frac{v}{r} \Hmunu \; ,\; \;  \partial_i \Hmunu = \frac{1}{r} \Hmunu \; , \; \\
       \partial_\alpha \hbarmunu &= \frac{v}{r} \hbarmunu  \;, \end{split}
\end{equation}
where Greek (Latin) indices run from $0,...,3$ ($1,...,3$). The same relations hold for the scalar field.\footnote{When applying the same procedure to the scalar field, there appears a subtlety due to the finite scalar mass, which introduces an additional scale. We will comment on this later in this section when constructing the scalar propagator.} Because of the different scaling of spatial and temporal derivatives, we further follow Refs.~\cite{Goldberger:2004jt,Goldberger:2007hy} and define the Fourier transformed fields
\begin{equation} \label{eq:graviton_fourier}
    \begin{split}
    \Hmunu (x) &= \int \frac{d^3\kbf}{(2\pi)^3} \exp\left(i \kbf \xbf\right) \Hkbfmunu (x^0) \; , \\
    \PhiBar(x) &= \int \frac{d^3\kbf}{(2\pi)^3} \exp(i\kbf \xbf) \PhiBar_\kbf(x^0)\; .  
    \end{split}
\end{equation} 
In this way, the $\sim 1/r$ fluctuations from the spatial momenta are disentangled from the fields.
Derivatives acting on the coordinates $\xbf^i$ can then directly be evaluated. Thus, all remaining derivatives acting on the fields scale identically,
\begin{equation}
    \partial_0 H_{\kbf \mu \nu} = \frac{v}{r} \Hmunu \; ,\;  \; \partial_\alpha \hbarmunu = \frac{v}{r} \hbarmunu \; .
\end{equation}\\

\paragraph*{\bf Power counting.} Having established the hierarchy of scales, we now introduce power counting rules. These are required to systematically collect all operators that contribute to the desired order in the PN expansion. First, we again consider the virial theorem to find
\begin{equation}
    \frac{M}{r \MPl^2} \sim v^2 \; .
\end{equation}
By introducing the angular momentum of the source $L = r M v$, we can conclude that 
\begin{equation}
    \frac{M}{\MPl} \sim (L v)^\frac{1}{2} \; .
\end{equation}
This factor appears, e.g., when considering the coupling of a graviton to the NS worldline.

Next, we consider the scaling of the graviton field by constructing the propagator from the quadratic term of the Einstein-Hilbert action \eqref{eq:SEH}. For the radiation modes this yields \cite{Goldberger:2004jt}
\begin{equation}
    \expval{T\Big(h_{\mu\nu} (x) h_{\alpha\beta} (y)\Big)} = D(x-y) P_{\mu\nu;\alpha\beta} \; ,
\end{equation}
with
\begin{equation} \label{eq:graviton_propagator_general}
    \begin{split}
        P_{\mu \nu; \alpha \beta} &= \frac{1}{2} \left( \eta_{\mu \alpha} \eta_{\nu \beta} + \eta_{\nu \alpha} \eta_{\mu \beta} - \eta_{\mu \nu} \eta_{\alpha \beta} \right) \; , \\
        D(x-y) &= \int \frac{d^4 k}{(2\pi)^4} \frac{i \exp(-ik(x-y))}{k^2 + i\epsilon} \; .
    \end{split}
\end{equation}
We can now simply apply the velocity dependence of the on-shell momenta and obtain
\begin{equation}
    \mu \nu \; 
    \begin{tikzpicture}
    \begin{feynman}
        \vertex (a);
        \vertex [right=2cm of a] (b);
        \diagram* {
          (a) -- [photon, very thick, momentum=\(k\)] (b) 
        };
    \end{feynman}
    \end{tikzpicture} \; \alpha \beta \;\sim\; \left(\frac{v}{r}\right)^4 \left(\frac{v}{r}\right)^{-2}  = \left(\frac{v}{r}\right)^{2} \; .
\end{equation}
Therefore, the soft graviton modes scale as
\begin{equation}
     \hbarmunu \sim \frac{v}{r} \; .
\end{equation}
In terms of the hard modes $H_{\kbf\mu\nu}$, the relevant part of the Lagrangian reads \cite{Goldberger:2004jt}
\begin{equation}
    \begin{split}
        \mathcal{L}_\mathrm{EH} \supset &-\frac{1}{2} \int \frac{d^3\kbf}{(2\pi)^3} \Big[\kbf^2 \Hkbfmunu H^{\mu\nu}_{-\kbf} - \frac{\kbf^2}{2} H_\kbf H_{-\kbf} \\ 
        &- \partial_0 \Hkbfmunu \partial_0 H^{\mu\nu}_{-\kbf} + \frac{1}{2} \partial_0 H_\kbf \partial_0 H_{-\kbf}  \Big] \; . 
    \end{split}
\end{equation}

We note that the second line is suppressed by the velocity since $k^0/|\kbf| \sim v \ll 1$. Therefore, these terms can be treated as a perturbation
\begin{equation} \label{eq:propagator_expansion}
    \expval{T\Big( H_{\kbf\mu\nu} (t_1) H_{\qbf\alpha\beta} (t_2) \Big)} \sim -\frac{i}{|\kbf|^2} \left(1 + \frac{(k^0)^2}{|\kbf|^2} + ...\right) \; .
\end{equation}
Hence, the leading order expression is given by
\begin{equation} \label{eq:potential_graviton_propagator}
\begin{split}
    \mu \nu \; 
    &\begin{tikzpicture}
    \begin{feynman}
        \vertex (a);
        \vertex [right=2cm of a] (b);
        \diagram* {
          (a) -- [potentialGraviton, thick] (b),
          (a) -- [scalar, thick, momentum=\(k\), opacity=0] (b)
        };
    \end{feynman}
    \end{tikzpicture} \; \alpha \beta \\
    &\qquad = -(2 \pi)^3 \frac{i}{|\kbf|^2} \delta^{(3)}(\kbf + \mathbf{q})  \delta(t_1 - t_2) P_{\mu \nu; \alpha \beta} \\
    &\qquad \sim  \left(\frac{1}{r}\right)^{-2} \left(\frac{1}{r}\right)^{-3} \left(\frac{r}{v}\right)^{-1} = v r^4  \; .
\end{split}
\end{equation}

\bgroup
\def\arraystretch{1.6}
\setlength\tabcolsep{7pt}
\begin{table}[]
    \centering
    \begin{tabular}{|c|c|c|c|c|c|}
        \hline
        \bf{Quantity} & $\displaystyle  M/\MPl$ & $\displaystyle   \hbarmunu$ & $\displaystyle \Hkbfmunu$ & $\displaystyle \Bar{\phi}$ & $\displaystyle \Phi_\kbf$  \\
        \hline
        \bf{Scaling} & $\displaystyle  (L v)^\frac{1}{2}$ & $\displaystyle v/r$ & $\displaystyle v^\frac{1}{2} r^2$ & $\displaystyle v/r$ & $\displaystyle v^\frac{1}{2} r^2$\\ 
        \hline 
    \end{tabular}
    \caption{Power counting rules for the binary EFT of a scalar-tensor theory.}
    \label{tab:power_counting}
\end{table} \noindent
\egroup
From this we read off the scaling of a potential graviton mode
\begin{equation}
    \Hkbfmunu \sim v^\frac{1}{2} r^2 \; .
\end{equation}
Equation~\eqref{eq:potential_graviton_propagator} corresponds to an instantaneous interaction. The higher-order terms of the expansion in Eq.~\eqref{eq:propagator_expansion} are incorporated as corrections to the propagator, diagrammatically displayed by
\begin{equation}
\begin{split}
    \mu \nu \; &
    \begin{tikzpicture}
    \begin{feynman}
        \vertex (a);
        \node [right=1cm of a, crossed dot] (b);
        \vertex [right=1.2cm of b] (f1);
        \diagram* {
          (a) -- [potentialGraviton, thick] (b) -- [potentialGraviton, thick] (f1)
        };
    \end{feynman}
    \end{tikzpicture} \; \alpha \beta\\
    &\quad = -(2 \pi)^3 \frac{i}{|\kbf|^4} \delta^{(3)}(\kbf + \mathbf{q}) \frac{\partial}{\partial t_1 \partial t_2} \delta(t_1 - t_2) P_{\mu \nu; \alpha \beta} \; .
\end{split}
\end{equation}

The above procedure is repeated for the scalar degree of freedom. From the quadratic term in Eq.~\eqref{eq:scalar_action} we obtain the well-known propagator of a massive scalar field
\begin{equation} \label{eq:massive_scalar_propagator}
    \begin{split}
        \expval{T\Big(\Phi(x) \Phi(y)\Big)}  &= \int \frac{d^4 k}{(2\pi)^4} \frac{i \exp(-ik(x-y))}{k^2 - m_s^2 + i\epsilon}  \; .
    \end{split}
\end{equation}
For the off-shell modes we neglect the $i\epsilon$ term and expand
\begin{equation}
    \label{eq:massive_scalar_propagator_expansion}
    \frac{i}{k^2 - m_s^2} = -\frac{i}{|\kbf|^2 + m_s^2} \left(1 + \frac{(k^0)^2}{|\kbf|^2 + m_s^2} + ...\right) \; .
\end{equation}
This gives us the Feynman rules for the momentum space propagator
\begin{equation}
    \begin{split}
    &\begin{tikzpicture}
    \begin{feynman}
        \vertex (a);
        \vertex [right=2cm of a] (b);
        \diagram* {
          (a) -- [potentialScalar, thick, momentum=\(k\)] (b)
        };
    \end{feynman}
    \end{tikzpicture}\\
     &\qquad \quad = \;-(2 \pi)^3 \frac{i}{|\kbf|^2 + m_s^2} \delta^{(3)}(\kbf + \mathbf{q})  \delta(t_1 - t_2) \; ,
     \end{split}
\end{equation}
as well as its LO correction
\begin{equation}
    \begin{split}
    & \begin{tikzpicture} 
    \begin{feynman}
        \vertex (a);
        \node [right=1cm of a, crossed dot] (b);
        \vertex [right=1.2cm of b] (f1);
        \diagram* {
          (a) -- [potentialScalar, thick] (b) -- [potentialScalar, thick] (f1)
        }; 
    \end{feynman}
    \end{tikzpicture} \\
    &\quad = -(2 \pi)^3 \frac{i}{(|\kbf|^2 + m_s^2)^2} \delta^{(3)}(\kbf + \mathbf{q}) \frac{\partial}{\partial t_1 \partial t_2} \delta(t_1 - t_2) \; .
    \end{split}
\end{equation}
In terms of the velocity powers, we obtain the same scaling relations as for the graviton case
\begin{equation}
    \PhiBar_\kbf \sim v^\frac{1}{2} r \; , \quad \phiBar \sim \frac{v}{r} \; .
\end{equation}
Table~\ref{tab:power_counting} contains a summary of the relevant power counting rules.

Please note that, when expanding Eq.~\eqref{eq:massive_scalar_propagator_expansion}, we implicitly assumed that the scalar mass is of similar order as $|\kbf|$, i.e.~$m_s \sim |\kbf| \sim 1/r$. However, we would like to model the inspiral phase as a whole and thus not only focus on the stage in which $m_s \sim 1/r$. In the full EFT spirit, one therefore has to make a distinction between different phases of the inspiral, and from that construct different EFTs. More precisely, three distinct cases emerge for the potential scalars:
\begin{itemize}
    \item $m_s \gg 1/r$: The mass is too large to impact the inspiral phase. Essentially, as soon as $m_s \gtrsim 1/R_s$, all potential effects of the scalar field would need to be integrated out and would then only appear via Wilson coefficients in the point particle action. The same can also be seen from Eq.~\eqref{eq:massive_scalar_propagator_expansion}, since the mass would suppress the propagator. In this case, one would construct the EFT by neglecting all potential scalar interactions.
    
    \item $m_s \sim 1/r$: The decomposition of the propagator in Eq.~\eqref{eq:massive_scalar_propagator_expansion} is valid, the power counting is manifest, and one would include potential scalars in the further construction of the EFT.
    \item $m_s \ll 1/r$: The scalar field behaves essentially massless on the orbital separation scale. This becomes apparent since one could now further expand the propagator in Eq.~\eqref{eq:massive_scalar_propagator_expansion} in terms of the scalar mass $m_s$, with the leading order term corresponding to a massless propagator.
\end{itemize}
However, we decide not to make this distinction and instead use the presented expansion of the scalar propagator for all of the above cases. This has the great advantage that we do not have to construct multiple EFTs and instead have one that is valid for all the above-described cases. 

At the same time, this carries the disadvantage that we lose the manifest power counting since we do not fix the scale of $m_s$. Instead, when inspecting a certain diagram, we assign the scale to $m_s$ which will result in the overall lowest scaling for the considered diagram, which effectively means choosing $m_s \sim 1/r$ for all potential scalars. In this way, we keep all the diagrams that are relevant for all three cases described above. If we wish to specialize to a certain case, such as taking the massless limit $m_s \rightarrow 0$ for a particular quantity, e.g., the binding energy, we can still do this at the end of the calculations. The same is true for the large mass limit $m_s \gg 1/r$. Then, scalar contributions are exponentially suppressed by terms of the form $\sim \exp(-m_s r)$ (cf.~Sec.~\ref{subsec::conservative_dynamics}).

Similar considerations apply to soft scalar modes. We will later see that LO scalar radiation only exists if $m_s < v/r$, which naturally sets the scale $m_s \sim v/r$ for on-shell scalar radiation (cf.~Sec.~\ref{subsec::radiative_dynamics}). 
Therefore, all our results are valid for generic masses $m_s$.

\section{Construction of the Scalar-Tensor EFT} \label{sec:scalar_tensor_eft}
To obtain an effective theory at the soft scale, we integrate out the short-wavelength fields, i.e., the potential scalar and graviton modes. This leaves us with the effective action
\begin{equation}
    \exp\left(i \Seff \right) = \int \mathcal{D}\Phi \; \mathcal{D} \Hmunu \exp\left(i \mathcal{S} \right) \; ,
\end{equation}
where $\mathcal{S}$ is the full action from Eq.~\eqref{eq:action}. In the following, we demonstrate the necessary steps to compute this quantity. 
First, we derive all relevant interactions at the desired order in the PN expansion by expanding the full action.
These are organized in Feynman diagrams, where potential modes only appear as internal lines, while radiation modes enter externally. 
Subsequently, we compute the energy loss of the binary system, which ultimately leads to a prediction of the gravitational waveform. In this work, we limit ourselves to the first order in the PN expansion, which corresponds to $\mathcal{O}(v^2)$. Higher-order effects on the waveform will be studied in future work.

\subsection{Action Expansion}
First, we expand the full action \eqref{eq:action} in order to determine the interactions between the binary system and the respective fields. \\
    
\paragraph*{\bf Pure GR.} We start by considering the pure GR part of point-particle action $\Spp$.
The first term in Eq.~\eqref{eq:Spp} expands as
\begin{equation} \label{eq:Spp_expansion}
\begin{split}
     \Spp \simeq &-M \int d\Bar{\tau} - \frac{M}{2\MPl} \int d\Bar{\tau} h_{\mu\nu} \frac{dx^\mu}{d\Bar{\tau}} \frac{dx^\nu}{d\Bar{\tau}} \\
     &+ \frac{M}{8\MPl^2} \int d\Bar{\tau} \left(h_{\mu\nu} \frac{dx^\mu}{d\Bar{\tau}} \frac{dx^\nu}{d\Bar{\tau}}\right)^2\; + ... \; ,
\end{split}
\end{equation}
where we have not explicitly written down the sum over the constituents. In addition, we have introduced
\begin{equation}
    d\Bar{\tau}^2 \equiv \eta_{\mu\nu} dx^\mu dx^\nu \; .     
\end{equation}
We now expand this expression to $\mathcal{O}(\vbf ^4)$, which yields 
\begin{align}
    \begin{split}
        d\Bar{\tau} &= dt \left(\eta_{\mu\nu} \frac{dx^\mu}{dt} \frac{dx^\nu}{dt}\right)^\frac{1}{2} \\
        &= dt \left(1 - \vbf^2 \right)^{1/2} \simeq dt \left(1 - \frac{1}{2} \vbf^2 - \frac{1}{8} \vbf^4\right) \; . 
    \end{split}
\end{align}
In addition, we have
\begin{equation}
    \frac{dx^\mu}{d\Bar{\tau}} = \frac{dx^\mu}{dt} \frac{1}{(1-\vbf^2)^\frac{1}{2}} \simeq \frac{dx^\mu}{dt} \left(1 + \frac{1}{2} \vbf^2 + \frac{3}{8} \vbf^4\right) \; ,
\end{equation}
with $\displaystyle dx^\mu/ dt = (1, \vbf)$. Plugging this into Eq.~\eqref{eq:Spp_expansion}, we obtain
\begin{widetext}
    \begin{equation}
        \label{eq:S_pp_full_expansion}
        \begin{split}
            \Spp^{(0)} &= -M \int dt \left(1 - \frac{1}{2} \vbf^2 - \frac{1}{8} \vbf^4\right) \; ,\\
            \Spp^{(1)} &= -\frac{M}{2\MPl} \int dt \biggl(h_{00} + 2 h_{0i} \vbf^i + h_{ij} \vbf^i \vbf^j + \frac{h_{00}}{2} \vbf^2 + h_{0i} \vbf^i \vbf^2 + \frac{h_{ij}}{2} \vbf^i \vbf^j \vbf^2 + \frac{3}{8} h_{00}\vbf^4 \biggr) \; ,\\
            \Spp^{(2)} &= \frac{M}{8\MPl^2} \int dt \biggl(h_{00}^2 + 4 h_{00} h_{0i} \vbf^i + 4(h_{0i} \vbf^i)^2 + 2 h_{00} h_{ij} \vbf^i \vbf^j  
            + \frac{3}{2} h_{00}^2 \vbf^2 + 4 h_{0i} h_{jk} \vbf^i \vbf^j \vbf^k + 6 h_{00} h_{0i} \vbf^i \vbf^2 \\ 
            &\hspace{5cm} + \left(h_{ij} \vbf^i \vbf^j \right)^2 + 6 \left(h_{0i} \vbf^i\right)^2 \vbf^2 + 3 h_{00} h_{ij} \vbf^i \vbf^j \vbf^2 + \frac{15}{8} h_{00} \vbf^4 \biggr) \; .
        \end{split}
    \end{equation}
\end{widetext} 
This expression contains all couplings between the NS worldlines and the graviton field. Together with the power counting rules for radiation/potential gravitons from Sec.~\ref{sec:EFT}, we can now pick up all terms relevant at 1PN. To demonstrate this procedure, one may, e.g., consider the first term in $\Spp^{(1)}$. For the interaction between the worldline and a potential graviton with polarization $H_{00}$, we obtain 
\begin{equation}
    \begin{split}
     S_{H_{00}-\mathrm{NS}} &= -\frac{M}{2 \MPl} \int dt\, \frac{d^3 \kbf}{(2 \pi)^3} \exp\left(i \kbf \xbf \right) H_{\kbf00} \\
    &\sim (M v r)^\frac{1}{2} v^\frac{1}{2} \frac{r}{v} \frac{1}{r^3} r^2 \sqrt{v} = \sqrt{Mvr} = \sqrt{L} \; ,
    \end{split}
\end{equation}
The corresponding Feynman rule for the $H_{00}$-$\mathrm{NS}$ vertex reads
\begin{equation}
    \includegraphics[valign=c]{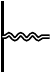} = -i \frac{M}{2 \MPl}  \int dt \int_\kbf \exp\left(i \kbf \xbf \right) \eta^{0\mu} \eta^{0\nu} \; .
\end{equation}
Note the factor of $i$ which enters via the expansion of the path integral.
Equivalently, for a radiation graviton we obtain the scaling
\begin{equation}
    S_{\Bar{h}_{00}-\mathrm{NS}} = \frac{M}{2 \MPl} \int dt \,\Bar{h}_{00} \sim (L v)^\frac{1}{2} \; .
\end{equation}
Regarding the Feynman rule, we have
\begin{equation}
    \includegraphics[valign=c]{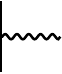} = -i \frac{M}{2 \MPl} \int dt  \; \Bar{h}_{00}\; .
\end{equation}
We repeat this procedure for all operators in Eq.~\eqref{eq:S_pp_full_expansion}. A summary of the relevant interactions, together with their Feynman rules and scaling relations, can be found in App.~\ref{app:feynman_rules}.\\
   
\paragraph*{\bf Einstein-Hilbert action.} 
In addition to the diagrams which arise from the coupling between the worldline and the gravitons, we need to take into account the graviton self-interactions which render $n$-graviton vertices. These contributions are collected by expanding the Einstein-Hilbert action $\SEH$. At cubic order, one, e.g., finds \cite{Goldberger:2007hy}
\begin{equation} 
    \begin{split}
    S_{H^3} &\sim \frac{1}{\MPl} \int dt (2\pi)^3 \delta^{(3)}\left(\sum_r \kbf_r\right) \kbf^2 \prod_i \frac{d^3 \kbf_r}{(2\pi)^3} H_{\kbf_r} \\
    &\sim \frac{v^2}{L^\frac{1}{2}}\; .
\end{split}
\end{equation}
Since this expression already scales as $\sim v^2$, we do not consider interaction terms of higher order. The corresponding momentum space Feynman rule reads \cite{Goldberger:2004jt}
\begin{equation}
    \begin{split}
    \includegraphics[valign=c]{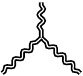} 
    &=-\frac{i}{4 \MPl} \delta(x_1^0-x_2^0) \delta(x_1^0-x_3^0) (2\pi)^3 \\
    &\qquad \quad \delta^{(3)}\left(\sum_r \kbf_r\right) \prod_r \frac{i}{\kbf_r^2} \times \sum_r \kbf_r^2 \; .
     \end{split}
\end{equation}\\

\begin{figure*}[]
    \centering
    \begin{subfigure}[b]{\textwidth}
    \includegraphics[]{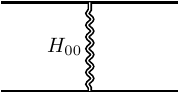}
    \includegraphics[]{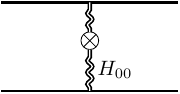}
    \includegraphics[]{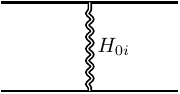}
    \includegraphics[]{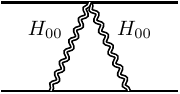}
    \includegraphics[]{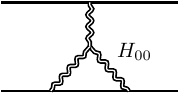}
    \subcaption{Binding energy diagrams at 1PN order in pure GR.}
    \label{fig:binding_energy:graviton}
    \end{subfigure}\vskip0.2cm
    \begin{subfigure}[b]{\textwidth}
    \includegraphics[]{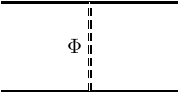}
    \includegraphics[]{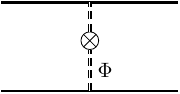}
    \includegraphics[]{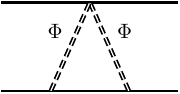}
    \includegraphics[]{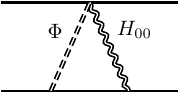}
    \includegraphics[]{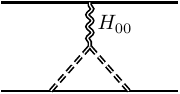}
    
    \vspace{0.2cm}
    
    \includegraphics[]{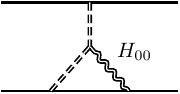}
    \includegraphics[]{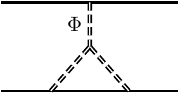}
    \subcaption{Scalar contribution to the binding energy.}
    \label{fig:binding_energy:scalar}
    \end{subfigure}
    \caption{All diagrams contributing to the binding energy at 1PN order.}
    \label{fig:binding_energy_diagrams}
\end{figure*}

\paragraph*{\bf Scalar contributions.}
Lastly, we consider the scalar sector. We first perform an expansion of the $\phi$-dependent terms in the point-particle action. From the linear term, we obtain all couplings between the worldline and one scalar
\begin{widetext}
        \begin{equation}
            \begin{split}
                S_\mathrm{pp\phi}^{(0)} &= -\frac{q_n}{\MPl} \int dt \left(1 - \frac{1}{2} \vbf^2 - \frac{1}{8} \vbf^4 \right) \phi \; ,\\
                S_\mathrm{pp\phi}^{(1)} &= -\frac{q_n}{2\MPl^2} \int dt \biggl(h_{00} + 2 h_{0i} \vbf^i + h_{ij} \vbf^i \vbf^j + \frac{h_{00}}{2} \vbf^2 + h_{0i} \vbf^i \vbf^2 + \frac{h_{ij}}{2} \vbf^i \vbf^j \vbf^2 + \frac{3}{8} h_{00}\vbf^4 \biggr) \phi \; , \\
                S_\mathrm{pp\phi}^{(2)} &= \frac{q_n}{8\MPl^3} \int dt \biggl(h_{00}^2 + 4 h_{00} h_{0i} \vbf^i + 4(h_{0i} \vbf^i)^2 + 2 h_{00} h_{ij} \vbf^i \vbf^j + \frac{3}{2} h_{00}^2 \vbf^2 + 4 h_{0i} h_{jk} \vbf^i \vbf^j \vbf^k + 6 h_{00} h_{0i} \vbf^i \vbf^2 \\ &\hspace{5.08cm} +\left(h_{ij} \vbf^i \vbf^j \right)^2 + 6 \left(h_{0i} \vbf^i\right)^2 \vbf^2 + 3 h_{00} h_{ij} \vbf^i \vbf^j \vbf^2 + \frac{15}{8} h_{00} \vbf^4 \biggr) \phi \; .
            \end{split}
        \end{equation}
This is repeated for the term in $\Spp$ which is quadratic in $\phi$, yielding
\begin{equation}
    \label{eq:S_phi1_full_expansion}
    \begin{split}
        S_\mathrm{pp\phi^2}^{(0)} &= -\frac{p_n}{\MPl^2} \int dt \left(1 - \frac{1}{2} \vbf^2  - \frac{1}{8} \vbf^4 \right) \phi^2 \; ,\\
        S_\mathrm{pp\phi^2}^{(1)} &= -\frac{p_n}{2\MPl^3} \int dt \biggl(h_{00} + 2 h_{0i} \vbf^i + h_{ij} \vbf^i \vbf^j + \frac{h_{00}}{2} \vbf^2 + h_{0i} \vbf^i \vbf^2 + \frac{h_{ij}}{2} \vbf^i \vbf^j \vbf^2 + \frac{3}{8} h_{00}\vbf^4 \biggr) \phi^2 \; , \\
        S_\mathrm{pp\phi^2}^{(2)} &= \frac{p_n}{8\MPl^4} \int dt \biggl(h_{00}^2 + 4 h_{00} h_{0i} \vbf^i + 4(h_{0i} \vbf^i)^2 + 2 h_{00} h_{ij} \vbf^i \vbf^j 
        + \frac{3}{2} h_{00}^2 \vbf^2 + 4 h_{0i} h_{jk} \vbf^i \vbf^j \vbf^k + 6 h_{00} h_{0i} \vbf^i \vbf^2 \\ 
        &\hspace{5.08cm} + \left(h_{ij} \vbf^i \vbf^j \right)^2 + 6 \left(h_{0i} \vbf^i\right)^2 \vbf^2 + 3 h_{00} h_{ij} \vbf^i \vbf^j \vbf^2 + \frac{15}{8} h_{00} \vbf^4 \biggr) \phi^2 \; .
        \end{split}
\end{equation}
\end{widetext}
\newpage
We can now derive, e.g., the velocity scaling of the interaction between a scalar potential mode and the worldine,
\begin{equation}
    S_{\PhiBar - \mathrm{NS}} \sim \frac{q_n}{\MPl} \int dt \int d^3 \kbf \, \PhiBar_\kbf \sim L^\frac{1}{2} \frac{q_n}{M_n} \; .
\end{equation}
Here $M_n$ and $q_n$ denote the mass and scalar charge of the $n^\mathrm{th}$ constituent. This results in the Feynman rule
\begin{equation}
    \includegraphics[valign=c]{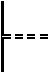} 
    =-i \frac{q_n}{\MPl} \int dt \int_{\kbf} \exp\left(i \kbf \xbf \right) \; . \\
\end{equation}
Similarly, the interaction between a soft mode and the worldline scales as
\begin{equation}
    S_{\phiBar-\mathrm{NS}} \sim \frac{q_n}{\MPl} \int dt \, \phiBar \sim (Lv)^\frac{1}{2} \frac{q_n}{M_n} \; .
\end{equation}
The Feynman rule reads
\begin{equation}
    \includegraphics[valign=c]{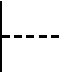} 
    =-i \frac{q_n}{\MPl} \int dt \,\phiBar \; . \\
\end{equation}
Note that the scalar-worldline interactions depend on the EFT parameters $q_n$ and $p_n$, which are in general model dependent. To remain model independent, we keep these coefficients generic throughout our entire computation. Our results may then be adapted to any theory by performing a matching procedure (see \cite{Sagunski:2017nzb,Huang:2018pbu,Zhang:2021mks} for LO examples). In Appendix~\ref{app:validity}, we comment on constraints on the EFT parameters to retain the validity of our treatment.

The remaining piece is the scalar action, which allows us to work out the interactions between the graviton and the scalar field, as well as $n$-scalar self-interactions. First, we expand the metric determinant
\begin{equation}
    \begin{split}
        \sqrt{-g} = 1 &+ \frac{1}{2\MPl} h_{\mu\nu} \eta^{\mu\nu} - \frac{1}{4 \MPl^2} h^{\mu\nu}  h_{\mu\nu} \\
        &+ \frac{1}{8 \MPl^2} (h_{\mu\nu} \eta^{\mu\nu})^2 + ... \; .
    \end{split}
\end{equation}
With the generic potential from Eq.~\eqref{eq:scalar_potential}, the potential part of the action then reads
\begin{equation} \label{eq:scalar_action_expansion}
    \begin{split}
        S_\phi \supset -\int &d^4x \biggl[\frac{m_s^2}{2} \phi^2 + \MPl \frac{c_3}{3!} \phi^3 + \frac{\lambda}{4!} \phi^4 
        \\
        &+ \frac{m_s^2}{4 \MPl} h_{\mu\nu} \eta^{\mu\nu}\phi^2
        + \frac{c_3}{2 \cdot 3!} h_{\mu\nu} \eta^{\mu\nu}\phi^3 \\
        &+ \frac{\lambda}{2\cdot 4! \MPl} h_{\mu\nu} \eta^{\mu\nu}\phi^4 + ... \biggr]  \; .
    \end{split}
\end{equation}
The kinetic term expands as 
\begin{equation}
    \begin{split}
        S_\phi \supset \int d^4x \biggl[ \frac{1}{2} \eta^{\mu\nu} &\partial_\mu \phi \partial_\nu \phi - \frac{h^{\mu\nu}}{2 {\MPl}} \partial_\mu \phi \partial_\nu \phi \\
        &+ \frac{h \eta^{\mu\nu}}{4 {\MPl}}  \partial_\mu \phi \partial_\nu \phi - \mathcal{O}(h^2)\biggr] \; ,
    \end{split}
\end{equation}
where $h = h_{\alpha \beta} \eta^{\alpha \beta}$. This renders the $h-\phi^2$ vertex
\begin{equation} \label{eq:hphiSq_vertex}
    \begin{split}
        S_{h\phi^2} \sim \frac{h^{\mu\nu}}{2 \MPl} k_\mu k_\nu \phi^2 &+ \frac{h \eta^{\mu\nu}}{4 \MPl} k_\mu k_\nu \phi^2 \\ 
        &- \frac{m_s^2}{4 \MPl} h^{\mu \nu} \eta_{\mu \nu} \phi^2 \; .
    \end{split}
\end{equation}
Considering only the momentum-dependent part, we have
\begin{equation}
    S_{h\phi^2} \sim \left(\frac{\eta^{\alpha\beta} P_{00;\alpha \beta}}{2} \eta_{\mu\nu} - P_{00;\mu\nu} \right) k^\mu k^\nu = \delta_{\mu 0} \delta_{\nu 0} k^\mu k^\nu \; .
\end{equation}
Hence, this expression only has a finite contribution for the time components of the four-momentum. This, however, is always suppressed relative to the mass vertex in Eq.~\eqref{eq:hphiSq_vertex}. Therefore, we can neglect the couplings generated by the kinetic term in the following, i.e.,~only the term $\sim m_s^2 h^{\mu \nu} \eta_{\mu\nu}$ remains. This is exactly the reason why the $h -\phi^2$ vertex vanishes in the massless limit \cite{Kuntz:2019zef} at the considered PN order.

From Eq.~\eqref{eq:scalar_action_expansion} we also obtain the scaling of the three-scalar vertex
\begin{equation}
    \begin{split}
         S_{\PhiBar^3} &\sim c_3 \MPl \int dt  \prod_r \int d^3\kbf_r\PhiBar_{\kbf_r} \delta^{(3)} \bigg(\sum_i \kbf_i \bigg) \\
         &\sim c_3 \MPl r v^\frac{1}{2} \sim c_3 \MPl^2 r^2 \frac{v^2}{L^\frac{1}{2}} \; .
    \end{split}    
\end{equation}
As for the graviton, we neglect higher orders since this expression already scales $\sim v^2$. The corresponding Feynman rule reads
\begin{equation}
    \begin{split}
    \includegraphics[valign=c]{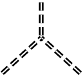}  
    &=-i \frac{c_3}{3!} \MPl \delta(x_1^0-x_2^0) \delta(x_1^0-x_3^0) \\
    &\qquad \quad \times (2\pi)^3 \delta^{(3)}\left(\sum_r \kbf_r\right) \prod_r \frac{i}{\kbf_r^2 + m^2}  \; .
    \end{split}
\end{equation}
The remaining scalar interactions and their associated Feynman rules are listed in Appendix~\ref{app:feynman_rules}. We now have all the ingredients to compute the Feynman diagrams which contribute to the binding energy, as well as the radiation power of the binary system. 

\bgroup
\setlength\cellspacetoplimit{4pt}
\setlength\cellspacebottomlimit{4pt}
\setlength\tabcolsep{5.8pt}
\begin{table*}[t]
  \centering
  \begin{tabular}{|c|c|c|c|c|c|Sc|c|c|c|c|}
  \hline
    $M$ & $x$ & $\gamma$ & $\mu$ & $\nu$ & $Q$ & $\Bar{Q}$ & $\displaystyle \Bar{m}_s$ & $\Bar{c}_3$ &$\xi_q$ & $\xi_p$ 
    \\ \hline
    \rule{0pt}{0.65cm}
    $\displaystyle M_1 + M_2 $ & $\displaystyle (G M \omega)^\frac{2}{3}$ & $ \displaystyle \frac{G M}{r}$ & $\displaystyle \frac{M_1 M_2}{M}$ & $\displaystyle \frac{\mu}{M}$ &  $\displaystyle \frac{q_1+q_2}{M}$ & $\displaystyle \frac{q_1 q_2}{M_1 M_2}$ & $\displaystyle G M m_s$ & $\displaystyle c_3  G M^2  Q  \Bar{Q}$& $ \displaystyle \frac{q_1^2 M_2 + q_2^2 M_1}{M M_1 M_2}$ & $\displaystyle \frac{p_1 q_2^2 + p_2 q_1^2}{M M_1 M_2}$ \rule[-0.5cm]{0pt}{0pt} \\ 
    \hline 
  \end{tabular}
  \newline
  \vspace{0.01cm}
  \newline
  \setlength\tabcolsep{4.5pt}
  \begin{tabular}{|c|c|c|c|c|c|Sc|c|}
  \hline \rule{0pt}{0.45cm}
    $v$ & $\delta q$ & $\delta M$ & $\Xi_p$ & $\Xi_c$ & $\Xi_q$ & $g_1$ & $g_2$ \\ \hline
    \rule{0pt}{0.65cm}
    $(G M f)^\frac{1}{3}$&$\displaystyle \frac{M_2 q_1-M_1 q_2}{M^2}$ & $ \displaystyle M_1 - M_2$ & $\displaystyle \frac{M_1 p_2 q_1 - M_2 p_1 q_2}{M^3} \delta q$ & $\displaystyle \frac{c_3 \Bar{Q} \delta M \delta q}{M}$ &  $\displaystyle \frac{M_2^2 q_1 + M_1^2 q_2}{M^3}$ & 
    $\displaystyle \frac{M_2^2 q_1 - M_1^2 q_2}{M^3}$ 
    & $\displaystyle \frac{M_2^3 q_1 - M_1^3 q_2}{M^4}$  \rule[-0.5cm]{0pt}{0pt}\\ \hline 
  \end{tabular}
  \caption{Overview of the parameters introduced in our work. Besides the typical PN expansion variables, we rescale the EFT parameters for convenience.}
  \label{tab:dimenionless_params}
\end{table*}
\egroup 

\subsection{Conservative Dynamics} \label{subsec::conservative_dynamics}
Let us first consider the orbital scale and compute the diagrams that contribute to the conservative dynamics of the system. To do so, we collect all diagrams with internal potential modes and no external legs. First, we derive the potential energy in pure GR, before including the effects from the scalar field. Eventually, we derive the scalar modifications to the Kepler relation and binding energy. \\

\paragraph*{\bf GR contribution.} Applying the Feynman rules from Appendix~\ref{app:feynman_rules}, we find five diagrams which scale $\sim v^2$ at most, hence contribute up to 1PN. These are depicted in Fig.~\ref{fig:binding_energy:graviton}. 
Please note that the first diagram, the exchange of a potential graviton with polarization $H_{00}$, contains only the LO term. For simplicity, we evaluate the third diagram involving a graviton with polarization $H_{0i}$ by taking into account higher-order velocity dependencies, such as $\sim H_{0i} \vbf^i$ or $\sim H_{00} \vbf^2$.

For the complete calculation of all the relevant processes at 1PN we refer the reader to Appendix~\ref{app:binding_energy_computation}. Here we simply state the final result. Including the kinetic term, we obtain 
\begin{equation}
    L_\mathrm{GR} = \frac{1}{2} \sumi M_i \vbf_i^2 + G \frac{M_1 M_2}{r} + \LEIH \; ,
\end{equation}
where $\LEIH$ is the Einstein-Infeld-Hoffmann Lagrangian \cite{Einstein:1938yz}
\begin{equation}
    \begin{split}
    \LEIH = &\frac{1}{8} \sumi M_i \vbf_i^4 + G \frac{M_1 M_2}{2r} \Biggr[3 (\vbf_1^2 + \vbf_2^2) 
    - 7 (\vbf_1 \cdot \vbf_2) \\
    &- \frac{(\vbf_1 \cdot \rbf)(\vbf_2 \cdot \rbf)}{r^2}\Biggr] - G^2 \frac{M_1 M_2(M_1 + M_2)}{2r^2} \; .
    \end{split}
\end{equation}\\

\paragraph*{\bf Scalar field contribution.}
Equivalently to the potential gravitons, the short-wavelength scalar modes mediate a force between the binary constituents. 
The contributions from the scalar field up to 1PN order are shown in Fig.~\ref{fig:binding_energy:scalar}. Note that these diagrams contain both pure scalar contributions, but also vertices at which scalars and gravitons couple together. The full computation of the diagrams is given in Appendix~\ref{app:binding_energy_computation}. Putting everything together, the scalar Lagrangian reads
\begin{widetext}
\begin{equation} \label{eq:scalar_lagrangian}
    \begin{split}
        L_{\phi} = &8 G q_1 q_2 \frac{e^{-m_s r}}{r} \bigg[1 - G\frac{M_1+ M_2}{r} - \frac{\vbf_1^2 + \vbf_2^2}{2} - \frac{(\vbf_1 \cdot \rbf_1)(\vbf_2 \cdot \rbf_2)}{2 \, r^2} (1 + m_s r) + \frac{\vbf_1 \cdot \vbf_2}{2}\bigg] \\
        &- 64 G^2 (p_1 q_2^2 + p_2 q_1^2) \frac{e^{-2 m_s r}}{r^2} + 4 m_s G^2 \frac{M_1 q_2^2 + M_2 q_1^2}{r} \bigg[e^{-2 m_s r} + 2m_s r {\rm Ei}(-2 m_s r) \bigg] \\
        &- 8 m_s G^2 q_1 q_2 \, \frac{M}{r} \, \bigg[ \log(2 m_s r) e^{-m_s r} - \Ei(-2 m_s r) e^{m_s r} \bigg] \\
        &+ G c_3 q_1 q_2 \frac{q_1 + q_2}{2 \pi m_s r} \bigg[ \Ei(-m_s r) e^{- m_s r} - \Ei(-3 m_s r) e^{m_s r} \bigg]\, ,
    \end{split}
\end{equation}
\end{widetext}
where $\Ei (x) = \int_{-\infty}^x dt \exp(t)/t$ denotes the exponential integral.

\begin{figure}[t]
    \centering
    \begin{subfigure}[b]{\textwidth}
    \includegraphics[]{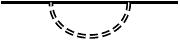}
    \hspace{0.6cm}
    \includegraphics[]{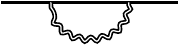}
    \caption{Diagrams contributing to the renormalization of the mass with no external vertices.}
    \label{renormalization::massrenorm}
    \end{subfigure}\vskip0.5cm
    \begin{subfigure}[b]{\textwidth}
    \includegraphics[]{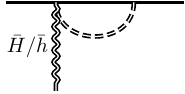}
    \hspace{0.6cm}
    \includegraphics[]{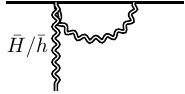}
    \subcaption{Diagrams contributing to the renormalization of the mass vertex with one external graviton.}
    \label{renormalization::singlegravitonvertex}
    \end{subfigure}\vskip0.5cm
    \begin{subfigure}[b]{\textwidth}
    \includegraphics[]{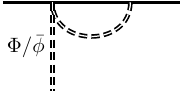}
    \hspace{0.6cm}
    \includegraphics[]{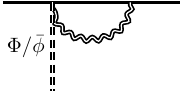}
    \subcaption{Diagrams contributing to the renormalization of the scalar charge vertex with one external scalar.}
    \label{renormalization::singlescalarvertex}
    \end{subfigure} 
    \caption{All self-force diagrams that contribute to the renormalization of the charge and mass up to 1PN order. The renormalization of the induced charge becomes relevant at higher PN orders.}
    \label{renormalization::1PN}
\end{figure}

In the above computations, we have not distinguished between the \textit{bare} and \textit{physical} masses and charges by neglecting the pure self-force diagrams shown in Fig.~\ref{renormalization::1PN}. In Appendix~\ref{app:renormalization} we demonstrate that the diagrams in Fig.~\ref{renormalization::massrenorm} can be absorbed by defining the physical mass
\begin{equation}
\label{renormalization::physicalMass_maintext}
    M_{\rm phys} = M_{\rm b} + \delta M_\text{sf} = M_{\rm b} + \frac{m_s\, q_\text{b}^2}{8 \pi \MPl^2} \; ,
\end{equation}
where $M_{\rm b}$ is the bare gravitational mass.
Similarly, the diagrams in Fig.~\ref{renormalization::singlescalarvertex} are taken into account by introducing the physical scalar charge
\begin{align}
\label{renormalization::physicalCharge_maintext}
    \begin{split}
        q_{\rm phys} &= q_\text{b} + \delta q_{\rm sf} + \frac{ q_\text{b}^2 c_3 \log 3}{16 \pi m_s} - \frac{m_s M_\text{b} q_\text{b} \,\gamma_E}{32 \pi \MPl^2} \\
        &= q_\text{b} + \frac{q_\text{b}}{4\pi} \bigg[ \frac{p_\text{b}\,m_s}{\MPl^2} + \frac{q_\text{b} \, c_3 \log 3}{4 \, m_s} - \frac{m_s M_\text{b}\,\gamma_E}{8 \MPl^2}\bigg] \;,
    \end{split}
\end{align}
with $q_{\rm b}$ denoting the bare scalar charge and $\gamma_E$ being the Euler-Mascheroni constant. Here the second term corresponds to the self-force diagram from Fig.~\ref{renormalization::singlescalarvertex}, while the two last terms arise when evaluating Fig.~\ref{fig:binding_energy:scalar}\,vi) and Fig.~\ref{fig:binding_energy:scalar}\,vii).
In the following, we will use the physical quantities, however, drop the corresponding subscript for brevity.\\

\paragraph*{\bf Modified Kepler relation.}
The total Lagrangian is given by 
\begin{equation} \label{eq:totalLagrangian}
    L_{\rm tot} = L_{\rm GR} + L_\phi \; ,  
\end{equation}
which allows us to compute the modification of the Kepler relation due to the scalar field.
In the remainder of this work, we specialize to circular orbits with
\begin{equation} \label{eq::circular_orbits}
        \rbf_1 = r_1 \begin{pmatrix}\cos \omega t\\ \sin \omega t \\ 0 \end{pmatrix} \; , \quad 
        \rbf_2 = r_2 \begin{pmatrix}\cos \left(\omega t +\pi\right) \\ \sin \left(\omega t+\pi \right) \\ 0 \end{pmatrix} \; .
\end{equation}
The Euler-Lagrange equations
\begin{equation}
    \frac{d}{dt} \frac{\partial L}{\partial \Dot{r}_i} - \frac{\partial L}{\partial r_i} = 0
\end{equation}
can now be used to derive relations between $r_1$, $r_2$ and $\omega$. Dotted quantities here denote derivatives with respect to time $t$. Note that since we consider circular orbits, we automatically have $\partial L / \partial \Dot{r}_i = 0$. In order to derive a relation on $r_1$ and $r_2$, it is convenient to consider
\begin{equation} \label{eq::r2_expansion}
    \frac{\partial L}{\partial r_1} - \frac{\partial L}{\partial r_2} = 0 \; .
\end{equation}
This reveals at 1PN order
\begin{align}
    r_2 = r \frac{M_1}{M} + \frac{\nu \delta M}{2} \left[ 8 G e^{-m_s r} \Bar{Q} + G - \frac{r^3 \omega^2}{M} \right]\;,
\end{align}
where $r = r_1 + r_2$ is the orbital separation. In order to obtain the above expression, we Taylor expanded in powers of $\Bar{Q} = q_1 q_2/(M_1 M_2)$ and only kept the LO term. Other terms, such as those proportional to $p_i$, cancel out at this order and thus do not contribute. To further obtain a relation between $r$ and $\omega$, i.e.,~the modified Kepler relation, we evaluate
\begin{equation} \label{eq:center_eq_2}
    \frac{\partial L}{\partial r_1} + \frac{\partial L}{\partial r_2} = 0 \; .
\end{equation}
After solving for $\omega$, we Taylor expand in $\Bar{Q}$ as well as in the other parameters describing the dynamics, e.g.,~$\xi_p$ (cf.~Table~\ref{tab:dimenionless_params}). 
In the following we first only focus on the LO contribution arising from the scalar field, i.e.,~the contribution proportional to $\Bar{Q}$, in order to illustrate the procedure. Subsequently, we present the full result.

At 1PN order and keeping only terms at most linear in $\Bar{Q}$ we obtain the modified Kepler relation
\begin{equation}
    \omega^2 = \frac{G M}{r^3} \bigg[ 1 + 8 \Bar{Q} (1 + m_s r) e^{-m_s r} + (\nu - 3) \frac{G M}{r} \bigg] \; .
\end{equation}
Rewriting this expression in terms of the usual PN expansion parameter, i.e.,~$\gamma = GM/r$ and $x = (G M \omega)^{2/3}$ (see~Table~\ref{tab:dimenionless_params}), we obtain the equivalent relation
\begin{equation}
    x^3 = \gamma^3 \bigg[ 1 + 8 \Bar{Q} \bigg(1 + \frac{\Bar{m}_s}{\gamma} \bigg) e^{-\Bar{m}_s / \gamma} + (\nu - 3) \gamma \bigg] \; .
\end{equation}
Inverting this equation for $\gamma$ and only keeping LO terms in $\Bar{Q}$, we obtain
\begin{align}
\label{inverse-kepler-law-leading-order}
    \begin{split}
        \gamma = \overbrace{x + \left(1 - \frac{\nu}{3}\right) x^2}^{\gamma_\text{GR} \equiv} - \Bar{Q} \frac{8}{3} (\Bar{m}_s + x) e^{-\Bar{m}_s / \gamma_\text{GR}}\;,
    \end{split}
\end{align}
where $\Bar{m}_s = G M m_s$. The parameter $\gamma_\text{GR}$ is the usual one which one would find in pure GR. It can in principle be exchanged with any higher-order accurate relation. Further, note that $e^{-\Bar{m}_s / \gamma_\text{GR}}$ does not admit a Taylor expansion at around $\gamma_\text{GR} = 0$. Instead, it is feasible to first expand $1/\gamma_\text{GR}$ and then expand the exponential function while leaving the LO term in the exponential, $-\Bar{m}_s/x$, untouched. More explicitly, first we expand
\begin{equation}
    -\frac{\Bar{m}_s}{\gamma_\text{GR}} = -\frac{\Bar{m}_s}{x} + \Bar{m}_s \left( 1 - \frac{\nu}{3} \right) + \Bar{m}_s \mathcal{O}(x) \; ,
\end{equation}
and thus
\begin{equation}
    e^{-\Bar{m}_s / \gamma_\text{GR}} = e^{-\Bar{m}_s / x} e^{\Bar{m}_s (1 - \nu/3) + \Bar{m}_s \mathcal{O}(x)} \;.
\end{equation}
If we assume $\Bar{m}_s$ to be on the hard scale, i.e.~$\Bar{m}_s \sim \gamma$, then we can directly Taylor expand around the LO term to obtain a well-behaved series, which can be truncated at the desired order. However, as we wish to keep the analysis more general, as to be applicable for any value of $\Bar{m}_s$, we are not directly in a position where we can do this expansion. Nonetheless, note that if one formally expands the second exponential, we obtain
\begin{equation}
\label{non-pert-expansion-exp}
    e^{-\Bar{m}_s / \gamma_\text{GR}} = e^{-\Bar{m}_s / x} \left[ 1 + \Bar{m}_s \left( 1 - \frac{\nu}{3} \right) + \mathcal{O}(\Bar{m}_s^2) \mathcal{O}(x) \right] \;.
\end{equation}
This expansion can generally be written as
\begin{equation}
    e^{-\Bar{m}_s / \gamma_\text{GR}} = \sum_{n=0}^\infty f_{n}(x) \frac{\Bar{m}_s^n}{n!} e^{-\Bar{m}_s / x}
\end{equation}
with $f_n(x)$ absorbing the remaining dependence on $x$ and some overall coefficients of order $\mathcal{O}(1)$\footnote{A more careful analysis reveals $f_n(x) = (1/x - 1/\gamma_\text{GR})^n$.}. Note that each term individually, if viewed as a function of $\Bar{m}_s$, reaches a maximum at $\Bar{m}_s = a x$, such that
\begin{equation}
    f_n(x) \frac{\Bar{m}_s^n}{n!} e^{-\Bar{m}_s / x} \leq f_n(x) x^n \frac{n^n}{n!} e^{-n} \leq f_n(x) x^n \;.
\end{equation}
Hence, a term proportional to $\Bar{m}_s^n$ can at most contribute with $x^n$. This means that effectively we can still assume $\Bar{m}_s \sim \gamma \sim x$ when doing this expansion, even if the scalar mass does not formally live on the hard scale. In other words, $\Bar{m}_s^n$ can never become more important than $x^n$.
By assuming $\Bar{m}_s \sim x$ we can make sure to always only keep those terms in the expansion that have a chance to become relevant at some point during the inspiral. The same reasoning also applies to the other functions that appear in the Lagrangian, such that we can also expand terms like $\Ei(-\Bar{m}_s/\gamma_\text{GR})$.

Continuing with the calculation and following the above discussion, we are allowed to truncate the expansion in Eq.~\eqref{non-pert-expansion-exp} to the desired PN order. At NLO, this is
\begin{equation}
\label{non-pert-expansion-exp2}
    e^{-\Bar{m}_s / \gamma_\text{GR}} = e^{-\Bar{m}_s / x} \left[ 1 + \Bar{m}_s \left( 1 - \frac{\nu}{3} \right) \right] \;,
\end{equation}
Inserting this relation into Eq.~\eqref{inverse-kepler-law-leading-order} and truncating at the relevant order again shows that the inverse Kepler law becomes
\begin{equation}
    \gamma = x + \left(1 - \frac{\nu}{3}\right) x^2 - \Bar{Q} \frac{8}{3} (\Bar{m}_s + x) e^{-\Bar{m}_s / x} \;.
\end{equation}
Note in this case the NLO term of Eq.~\eqref{non-pert-expansion-exp2} just gets truncated, so that the above-described expansion effectively results in replacing $\gamma_\text{GR}$ with $x$.

Having outlined the procedure, let us now give the full result that we use in the remainder of this work. Expanding consistently in all scalar parameters up to 1PN order, we obtain
\begin{widetext}
\begin{align}
    \begin{split}
    x^3 = &\frac{\gamma ^2 \bar{c}_3}{2 \pi \bar{m}_s} \left[e^{\frac{\bar{m}_s}{\gamma }} \left(\bar{m}_s-\gamma \right) \text{Ei}\left(-\frac{3 \bar{m}_s}{\gamma
   }\right)+ e^{-\frac{\bar{m}_s}{\gamma }} \left(\bar{m}_s+\gamma \right) \text{Ei}\left(-\frac{\bar{m}_s}{\gamma }\right)\right] -128 \gamma ^3 \xi _p e^{-\frac{2
   \bar{m}_s}{\gamma }} \left(\bar{m}_s+\gamma \right)\\
   &+4 \gamma ^3 \xi _q \bar{m}_s e^{-\frac{2 \bar{m}_s}{\gamma }} + 8 \gamma ^2 \bar{Q} e^{-\frac{\bar{m}_s}{\gamma }}
   \bigg[-\bar{m}_s e^{\frac{2 \bar{m}_s}{\gamma }} \left(\bar{m}_s-\gamma \right) \text{Ei}\left(-\frac{2 \bar{m}_s}{\gamma }\right)+\gamma  (3 \nu -4) \bar{m}_s \\
   &- \bar{m}_s \left(\bar{m}_s+\gamma \right) \log \left(\frac{2 \bar{m}_s}{\gamma }\right)+\bar{m}_s+2 \gamma ^2 (\nu -2)+\gamma \bigg] +\underbrace{\gamma ^3 (\gamma  (\nu -3)+1)}_{\displaystyle \equiv x_\mathrm{GR}} \; .
    \end{split}
    \label{eq:xPN_ST}
\end{align}
We invert the above expression to arrive at the inverse Kepler law
\begin{align} \label{eq:gamma_ST_1PN}
    \begin{split}
        \gamma = &\frac{\bar{c}_3}{6 \pi  \bar{m}_s} \left[-e^{\frac{\bar{m}_s}{x}} \left(\bar{m}_s-x\right) \text{Ei}\left(-\frac{3 \bar{m}_s}{x}\right)-  e^{-\frac{\bar{m}_s}{x}}  \left(\bar{m}_s+x\right)
   \text{Ei}\left(-\frac{\bar{m}_s}{x}\right)\right] + \frac{128}{3} x \xi _p e^{-\frac{2
   \bar{m}_s}{x}} \left(\bar{m}_s+x\right) \\
   &-\frac{4}{3} x \xi _q \bar{m}_s e^{-\frac{2 \bar{m}_s}{x}} + \frac{8}{9} \bar{Q} e^{-\frac{\bar{m}_s}{x}} \bigg[3 \bar{m}_s e^{\frac{2 \bar{m}_s}{x}}
   \left(\bar{m}_s-x\right) \text{Ei}\left(-\frac{2 \bar{m}_s}{x}\right)+\nu  \left(-4 x \bar{m}_s+\bar{m}_s^2-x^2\right) \\
   &-3 \left(x
   \bar{m}_s+\bar{m}_s^2+\bar{m}_s+x^2+x\right)+3 \bar{m}_s \left(\bar{m}_s+x\right) \log \left(\frac{2 \bar{m}_s}{x}\right)\bigg] \underbrace{- \frac{1}{3} (\nu - 3) x^2+x}_{\displaystyle  \equiv \gamma_\mathrm{GR}} \; .
    \end{split}
\end{align}
\end{widetext}
For this and all of the following results, we keep only terms up to linear order in the scalar parameters.

\paragraph*{\bf Binding energy.} The last quantity we compute regarding the conservative dynamics is the binding energy for a given orbital separation and frequency. Applying a Legendre transformation to the Lagrangian, we obtain the Hamiltonian $H$ for this system:
\begin{equation}
    H = \sum_{i= 1,2} \frac{\partial L}{\partial \Dot{\mathbf{x}}_i^k} \Dot{\mathbf{x}}_i^k - L \;.
\end{equation}
For the full Lagrangian given in Eq.~\eqref{eq:totalLagrangian}, we find the binding energy as a function of $\gamma$,
\begin{widetext}
\begin{align}
    \begin{split}
        \frac{E}{\mu} = &\frac{\bar{c}_3}{4 \pi  \bar{m}_s} \left[e^{-\frac{\bar{m}_s}{\gamma }} \left(\bar{m}_s-\gamma \right) \text{Ei}\left(-\frac{\bar{m}_s}{\gamma }\right)+e^{\frac{\bar{m}_s}{\gamma }}
   \left(\bar{m}_s+\gamma \right) \text{Ei}\left(-\frac{3 \bar{m}_s}{\gamma }\right)\right] -64 \gamma  \xi _p \bar{m}_s e^{-\frac{2 \bar{m}_s}{\gamma
   }} \\
   &+ \xi _q \left[-8 \bar{m}_s^2 \text{Ei}\left(-\frac{2
   \bar{m}_s}{\gamma }\right)-2 \gamma  \bar{m}_s e^{-\frac{2 \bar{m}_s}{\gamma }}\right] + \bar{Q} e^{-\frac{\bar{m}_s}{\gamma }} \bigg[-4 \bar{m}_s e^{\frac{2
   \bar{m}_s}{\gamma }} \left(\bar{m}_s+\gamma \right) \text{Ei}\left(-\frac{2 \bar{m}_s}{\gamma }\right) \\
   &+2 (- \nu \gamma +\gamma +2) \bar{m}_s + 4 \bar{m}_s \left(\gamma
   -\bar{m}_s\right) \log \left(\frac{2 \bar{m}_s}{\gamma }\right)-2 \gamma  (\gamma  (\nu -3)+2)\bigg] - \frac{1}{8} \gamma  (\gamma  (\nu -7)+4) \; .
    \end{split}
\end{align}
Plugging in Eq.~\eqref{eq:gamma_ST_1PN}, we obtain an expression depending on $x$:
\begin{align} \label{eq:binding_energy_full_1PN}
    \begin{split}
        \frac{E}{\mu} = &\frac{\bar{c}_3}{6 \pi  \bar{m}_s}  e^{-\frac{\bar{m}_s}{x}} \left[\left(2 \bar{m}_s-x\right) \text{Ei}\left(-\frac{\bar{m}_s}{x}\right)+e^{\frac{2 \bar{m}_s}{x}} \left(2 \bar{m}_s+x\right)
   \text{Ei}\left(-\frac{3 \bar{m}_s}{x}\right)\right] -\frac{64}{3} x \xi _p e^{-\frac{2 \bar{m}_s}{x}} \left(4 \bar{m}_s+x\right) \\
   &+\frac{4}{3} \xi _q \bar{m}_s \left[x \left(-e^{-\frac{2 \bar{m}_s}{x}}\right)-6 \bar{m}_s
   \text{Ei}\left(-\frac{2 \bar{m}_s}{x}\right)\right] +\frac{4}{9} \bar{Q} e^{-\frac{\bar{m}_s}{x}} \bigg[-6 \bar{m}_s e^{\frac{2 \bar{m}_s}{x}} \left(2 \bar{m}_s+x\right)
   \text{Ei}\left(-\frac{2 \bar{m}_s}{x}\right)  \\
   &-4 (\nu -3) \bar{m}_s^2 +4 ((\nu -3) x+3) \bar{m}_s+6 \bar{m}_s \left(x-2 \bar{m}_s\right) \log \left(\frac{2 \bar{m}_s}{x}\right) +x ((\nu -3) x-6)\bigg] +\frac{\nu  x^2}{24}+\frac{3 x^2}{8}-\frac{x}{2} \; .
    \end{split}
\end{align}
\end{widetext}
The definitions of all redefined parameters are collected in Table~\ref{tab:dimenionless_params}.

\begin{figure*}
    \begin{subfigure}{0.49\textwidth}
    \includegraphics[scale=0.87]{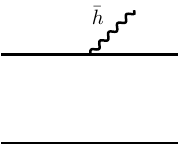}
    \hspace{0.17cm}
    \includegraphics[scale=0.87]{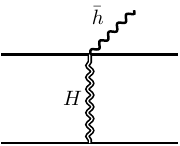}
    \hspace{0.17cm}
    \includegraphics[scale=0.87]{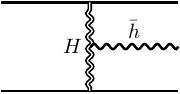}
    \subcaption{Pure GR radiation diagrams.}
    \label{fig::pure_graviton_emission}
    \end{subfigure}
    \begin{subfigure}{0.49\textwidth}
    \includegraphics[scale=0.87]{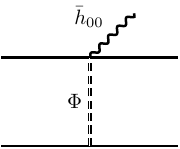}
    \hspace{0.17cm}
    \includegraphics[scale=0.87]{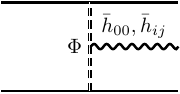}
    \subcaption{Modified graviton emission.}
    \label{fig::modified_graviton_emission}
    \end{subfigure}
    \begin{subfigure}{0.49\textwidth}
    \includegraphics[scale=0.87]{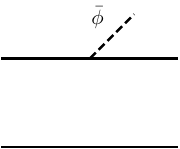}
    \hspace{0.17cm}
    \includegraphics[scale=0.87]{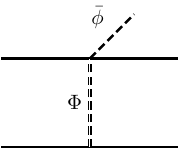}
    \hspace{0.17cm}
    \includegraphics[scale=0.87]{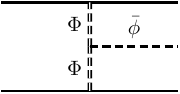}
    \subcaption{Pure scalar emission.}
    \label{fig::pure_scalar_emission}
    \end{subfigure}
    \begin{subfigure}{0.49\textwidth}
    \includegraphics[scale=0.87]{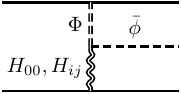}
    \hspace{0.17cm}
    \includegraphics[scale=0.87]{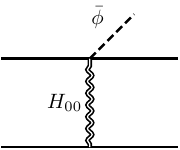}
    \subcaption{Modified scalar emission.}
    \label{fig::modified_scalar_emission}
    \end{subfigure}
    \caption{All emission diagrams appearing at 1PN, except for the last diagram in Fig.~\ref{fig::modified_scalar_emission}. Figure \ref{fig::pure_graviton_emission} shows all emission diagrams appearing in pure GR, while Fig.~\ref{fig::modified_graviton_emission} displays the new graviton emission diagrams appearing due to couplings with the scalar field. Similarly, Fig.~\ref{fig::pure_scalar_emission} shows pure scalar emission diagrams. Fig.~\ref{fig::modified_scalar_emission} shows all scalar emission diagrams that also contain graviton interactions.}
    \label{fig::all_emission_diagrams}
\end{figure*}

\subsection{Radiation} \label{subsec::radiative_dynamics}
Let us now discuss the computations involving soft degrees of freedom. Figure~\ref{fig::all_emission_diagrams} shows all emission diagrams that contribute up to 1PN order. This comprises pure GR, pure scalar, as well as mixed contributions. 
We do not carry out the full calculations of all pure GR diagrams. Instead, it is more sensible to only calculate the corrections due to the scalar field.
We then add these contributions in the fashion of a Taylor expansion to an existing template at high PN order in pure GR. 

\begin{figure}[t]
    \centering
    \includegraphics[]{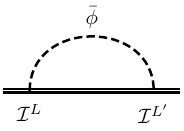}
    \hspace{0.8cm}
    \includegraphics[]{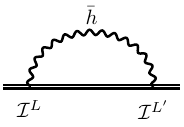}
    \caption{Self-force diagrams contributing to the LO radiation power starting at $-1$PN (left) and 0PN (right), respectively.}
    \label{fig::scalar_radiation_self_force}
\end{figure}

In the following, we first outline how the power loss via GWs is affected by the scalar contribution. Subsequently, the power loss via scalar radiation is computed. \\

\paragraph*{\bf Gravitational radiation.} After integrating out the orbital scale, the radiation gravitons are sourced by an effective energy-momentum tensor $T_{\rm eff}^{\mu\nu}$, which enters the effective action via
\begin{equation}
    S_{\rm eff} \supset S_{\text{src},\, \Bar{h}} = - \frac{1}{2 \MPl} \int dx^4\, T_{\rm eff}^{\mu\nu}(x) \Bar{h}_{\mu\nu}(x) \; .
\end{equation}
The general strategy to retain a consistent power counting is to expand $h_{\mu\nu}$ in multipole moments \cite{Ross:2012fc} around the source. We intend to compute the gravitational phase at NLO with regard to the scalar emission. As discussed in the subsequent section, the LO multipole in the scalar sector is given by the dipole, contributing at $-1$PN order. 
This implies it is sufficient to only consider the scalar effects on the quadrupolar radiation in the graviton sector. At LO, the quadrupole moment reads
\begin{equation}
    I^{ij} = \int d^3 \mathbf{x} \, T^{00} \left[\mathbf{x}^i \mathbf{x}^j\right]^\mathrm{STF} \; ,
\end{equation}
where STF denotes the symmetric and trace-free projection.
We find that none of the diagrams in Fig.~\ref{fig::modified_graviton_emission} contribute to this order. Therefore, the quadrupole moment takes the same form as in pure GR.
Explicitly, we have
\begin{equation}
    I^{ij} = \sum_{a = 1,2} M_a [\mathbf{r}_a^i \mathbf{r}_a^j]^\text{STF} \; .
\end{equation}
Using our expression for $\mathbf{r}_i$ and the expansion for $r_2$, Eqs.~\eqref{eq::circular_orbits} and~\eqref{eq::r2_expansion}, we further obtain
\begin{equation}
    I^{ij} = \mu \, [\mathbf{r}^i \mathbf{r}^j]^\text{STF} \; ,
\end{equation}
with $\mathbf{r} = \mathbf{r}_1 - \mathbf{r}_2$. Here we have already dropped all terms that will only contribute beyond the LO energy loss. Especially, all corrections due to the scalar field on $r_2$ will only contribute beyond 0PN and are thus negligible for the energy loss due to graviton emission at LO.
The energy loss is then given by the usual quadrupole formula
\begin{equation} \label{eq:power_loss_GR}
    \begin{split}
        P_{\Bar{h}} &= \frac{G}{5} \left< \left( \partial_t^3 I^{ij} \right)^2 \right> \\
        &= \frac{32 G}{5} M^2 \nu^2 r^4 \omega^6 \\
        &= \frac{32}{5 G} \nu^2 x^5 + \frac{1024}{15G} \nu^2 \Qbar e^{-\msbar / x} (\msbar + x) x^4 \; ,
    \end{split}
\end{equation}
where $\left<\dots\right>$ denotes time averaging. In the last line, we have dropped all terms which are of quadratic or higher power in $\Qbar$, and also those terms that would only contribute from 1PN order on. The first term is the usual pure GR quadrupole radiation, while the second term describes the modifications arising due to the scalar field. Notably, the presence of the scalar field affects the LO graviton radiation only via the modified Kepler relation, Eq.~\eqref{eq:xPN_ST}.\\

\paragraph*{\bf Scalar radiation.} 
Similar to the gravitational radiation, we obtain the energy loss by scalar emission by first identifying the source term that produces the radiation at linear order
\begin{equation}
    S_{\rm eff} \supset S_{\text{src},\, \phiBar} = \frac{1}{\MPl} \int d^4 x J(x) \phiBar(x) \; .
\end{equation}
All relevant scalar radiation diagrams contributing to $J$ are shown in Figs.~\ref{fig::pure_scalar_emission} and \ref{fig::modified_scalar_emission}. We follow Ref.~\cite{Ross:2012fc} to obtain the energy loss. First, we Taylor expand the field around the center of mass. This allows us to rewrite the source term via
\begin{equation}
    S_\text{src} = \frac{1}{\MPl} \int dt \sum_{l=0}^\infty \frac{1}{l!} \mathcal{I}^L \partial_L \phiBar \; ,
\end{equation}
where the derivatives $\partial_L \phiBar$ are  evaluated at $(t, \mathbf{x} = 0)$. $L$ denotes a multi-index, such that $x^L = x^{i_1}x^{i_2} \dots x^{i_l}$ and
\begin{equation}\label{eq::multipole_moments}
    \begin{split}
        \mathcal{I}^L = \sum_{p = 0}^\infty &\frac{(2l + 1)!!}{(2p)!! (2l + 2p + 1)!!} \\
        &\qquad \int d^3 \mathbf{x}\,r^{2p} \mathbf{x}_\text{STF}^L \,(\partial_t^2 + m_s^2)^{p} J \; ,
    \end{split}
\end{equation}
with $r = |\mathbf{x}|$, are the multipole moments of the source. A more detailed derivation of the above expression is given in Appendix~\ref{app:energy_loss_scalar}. Please note that Eq.~\eqref{eq::multipole_moments} deviates from Ref.~\cite{Ross:2012fc} since we consider a non-vanishing scalar potential.

Note further that the contributions to the source term arising from diagrams in which the radiating field couples directly to one of the worldlines can be calculated without performing the multipole expansion. The source term then always contains a $\delta$-function, which sets the position to the current position of the object whose worldline is being coupled to. The diagrams in which the radiating scalar instead couples to an interaction vertex whose position is integrated over, such as in Figs.~\ref{fig::pure_scalar_emission}\,iii) and \ref{fig::modified_scalar_emission}\,i), are slightly more complicated. They typically require performing the multipole expansion before the diagram can be solved analytically. In particular, this means that each term in the multipole expansion has to be calculated separately. In previous computations \cite{Huang:2018pbu,Bhattacharyya:2023kbh} this subtlety has not been taken into account. Instead, higher-order moments were calculated by supplementing the monopole term with an appropriate $\delta$-function in order to isolate the corresponding source term. 
This approach however does not capture the full dynamics, which can be verified by a direct computation of the higher-order moments. 
Therefore the computations in Refs.~\cite{Huang:2018pbu,Bhattacharyya:2023kbh} are incomplete, and we find additional terms contributing to the power loss via scalar radiation.
The explicit computations are relegated to Appendix~\ref{app:scalar_radiation_calculation}.

To derive the energy loss, we need to compute the imaginary part of the self-force diagrams in which the scalar field couples to the multipole moments. For our desired accuracy, it is sufficient to consider the LO topology depicted by the left diagram in Fig.~\ref{fig::scalar_radiation_self_force}. The calculation---see Appendix~\ref{app:energy_loss_scalar} for details---yields
\begin{align} \label{eq::scalar_radiation_generic_multipole}
    \begin{split}
        P_{\Bar{\phi}} = \frac{1}{4\pi^2 T} &\sum_{l=0}^\infty \frac{1}{l!(2l+1)!!}  \\
        &\int_{m_s}^\infty d\omega \, \omega \left( \omega^2 - m_s^2\right)^{l + 1/2} \left| \Tilde{\mathcal{I}}^L (\omega) \right|^2 \; ,
    \end{split}
\end{align}
where $T = 2\pi \delta(0)$. A tilde denotes Fourier-transformed quantities, e.g.~in this case
\begin{equation}
    \Tilde{\mathcal{I}}^L (\omega) = \int dt\,e^{i \omega t} \mathcal{I}^L (t).
\end{equation}
We decompose the total energy loss into its contributions from different multipole moments. Up to $0$PN order we obtain
\begin{equation} \label{eq::power_loss}
    \begin{split}
        P_{\Bar{\phi}}^\text{0PN} &= P_{\Bar{\phi}}^{l = 1} \left( 1 - \frac{m_s^2}{\omega^2} \right)^{3/2} + P_{\Bar{\phi}}^{l = 2} \left( 1 - \frac{m_s^2}{4\,\omega^2} \right)^{5/2} \; ,
    \end{split}
\end{equation}
where we factored out the common factor that inhibits scalar radiation for frequencies of $\omega < m_s$ ($\omega < m_s /2$) in the case of dipole (quadrupole) radiation. 
Note that no monopole radiation is emitted, since, as mentioned above, we only consider circular orbits.

\begin{figure*}[t]
    \centering
    \includegraphics[width=\textwidth]{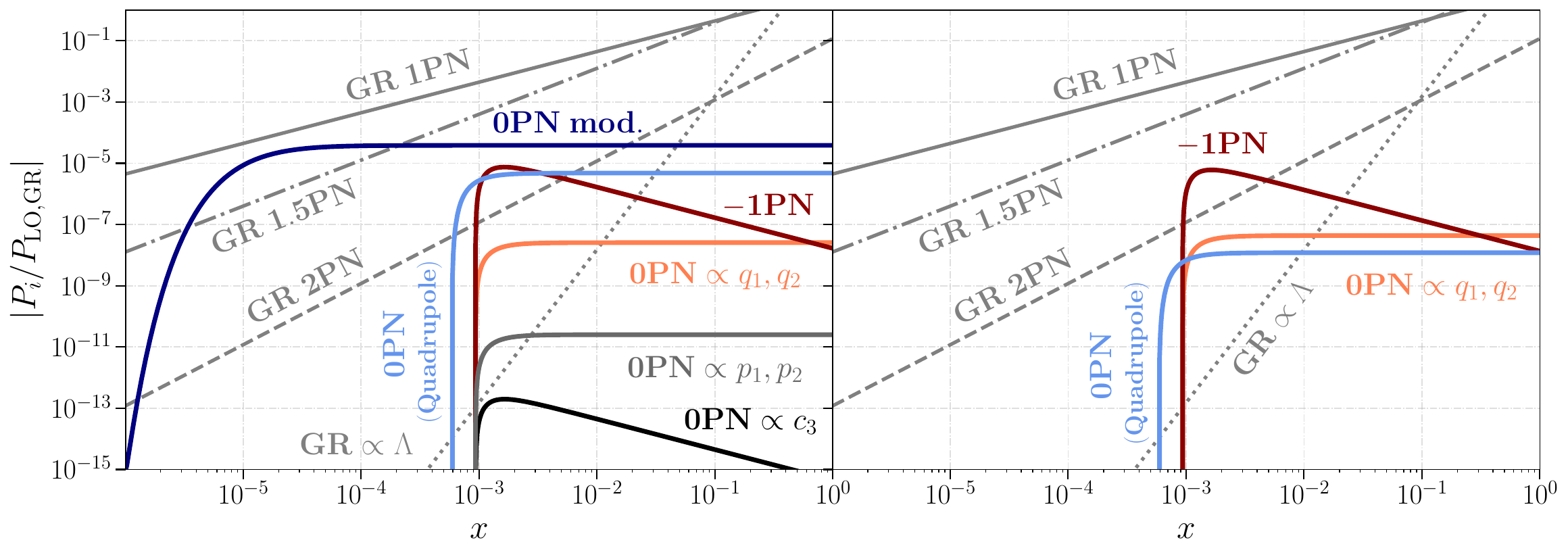}
    \caption{Power loss of an inspiraling binary system in massive scalar-tensor gravity, relative to the LO power loss in pure GR. Different colors indicate different contributions. We display the power loss arising due to scalar effects on the Kepler relation (dark blue, Eq.~\eqref{eq:power_loss_GR}), LO scalar dipole radiation (dark red, Eq.~\eqref{eq:scalar_power_loss:LO_dipole_x}), dipole effects due to the induced scalar charges (dark gray, Eq.~\eqref{eq:scalar_power_loss:p_dipole_x}) and cubic self-coupling (black, Eq.~\eqref{eq:scalar_power_loss:c3_dipole_x}), NLO scalar dipole radiation (orange, Eq.~\eqref{eq:scalar_power_loss:NLO_dipole_x}), as well as scalar quadrupole radiation (light blue, Eq.~\eqref{eq:scalar_power_loss:LO_quadropole_x}). The light gray curves indicate higher-order pure GR corrections up to 2PN order. In addition, we show typical pure GR finite-size contributions, where for the NS tidal deformability we choose a benchmark value of $\Lambda = 100$. In both panels, we fix $M_1 = 1.8 \,M_\odot$ and $M_2 = 2 \,M_\odot$, while the  employed model parameters are found in Table~\ref{tab::power_loss_plot_params}. The left panel corresponds to a typical NS-NS binary system. In the right panel, only one constituent carries a scalar charge---a constellation corresponding to a NS-BH binary. As a consequence, at the here considered order, the scalar degree of freedom has no effect on the modified Kepler relation. In addition, the contributions from the induced scalar charges $p_1$, $p_2$ and the self-coupling $c_3$ vanish. Also, please note that although we plot the power loss up to $x = 1$, our perturbative approach breaks down for large $x\lesssim 1$, i.e., our results can no longer be trusted. In this regime, other methods such as numerical relativity would have to be invoked to reliably describe the dynamics.}
    \label{fig:power_loss_plot}
\end{figure*}
\bgroup
\renewcommand{\arraystretch}{1.4}
\setlength\cellspacetoplimit{4pt}
\setlength\cellspacebottomlimit{4pt}
\setlength\tabcolsep{2.3pt}
\begin{table}[t]
  \centering
  \resizebox{\textwidth}{!}{\begin{tabular}{|c|c|c|c|c|c|c|}
  \hline
    & $m_s\,[\mathrm{eV}]$ & $q_1\,[M_1]$ & $q_2\,[M_2]$ & $p_1\,[q_1]$ & $p_2\,[q_2]$ & $c_3$ \\
    \hline 
    \text{Left}& $10^{-15}$& $1.8 \times 10^{-3}$& $2 \times 10^{-3}$& $5\times10^{-2}$& $5\times10^{-2}$ & $m_s^2/\MPl^2$\\
    \hline 
    \text{Right}& $10^{-15}$ & $1.8 \times 10^{-4}$ &$0$ & $5\times 10^{-3}$ &$0$ & $m_s^2/\MPl^2$ \\
    \hline
  \end{tabular}}
  \caption{Model parameters we employ in the respective panels of Fig.~\ref{fig:power_loss_plot}. The scalar mass corresponds to a characteristic value, which is typically not excluded by observations; see, e.g.,~the discussion in Ref.~\cite{Ramazanoglu:2016kul}. Also, the values for the scalar charges \cite{Creci:2023cfx,Huang:2018pbu} are typical values that lie in the observationally viable region. The value of the cubic self-coupling $c_3$ is motivated by $R^2$ gravity (cf.~Appendix~\ref{app:validity}).
  }\label{tab::power_loss_plot_params}
\end{table}
\egroup
When evaluating Eq.~\eqref{eq::scalar_radiation_generic_multipole}, we calculate the $p=0$ component for each diagram and only keep the $p=1$ contribution for the LO scalar emission diagram, i.e.,~the one in Fig.~\ref{fig::pure_scalar_emission}\,i). 
Judging from the power counting rules, one does not expect other diagrams to contribute for $l=p=1$ at the considered order.
However, a direct computation of the $l=p=1$ component of the diagram in Fig.~\ref{fig::modified_scalar_emission}\,i) reveals IR divergent terms in the limit of $m_s \rightarrow 0$.
Since in the massless limit the diagram itself vanishes as there is no corresponding operator in the action, these contributions are considered unphysical. At the here considered order, these divergent terms cancel out when calculating the power loss, and only a divergent term proportional to $\mathcal{O}(\Xi_c^2)$ remains. 
We suspect that these divergent terms are canceled by higher-order diagrams and postpone a more detailed study to future research.

Plugging the results from Appendix~\ref{app:scalar_radiation_calculation} into Eq.~\eqref{eq::power_loss}, we obtain
\begin{subequations} \label{eq:scalar_power_loss}
    \begin{align} 
        \begin{split} \label{eq:scalar_power_loss:LO_dipole}
            P_{\Bar{\phi}}^{l=1, \text{LO}} &= \frac{8}{3} G M^2 \delta q^2 r^2 \omega^4 ,
        \end{split}\\[2ex]
        \begin{split}\label{eq:scalar_power_loss:p_dipole}
            P_{\Bar{\phi}}^{l=1, p_1, p_2} &= \frac{256}{3} G^2 M^3 \Xi_p \, r \omega^4 ,
        \end{split}\\[2ex]
        \begin{split} \label{eq:scalar_power_loss:c3_dipole}
            P_{\Bar{\phi}}^{l=1, c_3} &= \frac{G M_1 M_2 M}{3 \pi m_s} \Xi_c (m_sr - 1) r^2 \omega^4 ,
        \end{split}\\[2ex]
        \begin{split} \label{eq:scalar_power_loss:NLO_dipole}
            P_{\Bar{\phi}}^{l=1, \text{NLO}} &= \frac{8 G M^2}{15} \delta q \bigg[ - 10 g_1 G M \\
            &{}\hspace{1cm} + g_2 r^3 ( m_s^2 - 6 \omega^2) \bigg] r \omega^4 ,
        \end{split}\\[2ex]
        \begin{split} \label{eq:scalar_power_loss:LO_quadropole}
            P_{\Bar{\phi}}^{l=2} &= \frac{128}{15} G M^2 \Xi_q^2 r^4 \omega^6 .
        \end{split}
    \end{align}
\end{subequations}
Here we have expanded the energy loss terms to only keep relevant contributions. More precisely, Eq.~\eqref{eq:scalar_power_loss:LO_dipole} contains the LO energy loss, which in our case is the usual dipole radiation term, formally entering at $-1$PN order. Likewise, Eq.~\eqref{eq:scalar_power_loss:NLO_dipole} contains the dipole radiation terms which enter at 0PN order. The terms which contain effects due to $p_1$, $p_2$ (Eq.~\eqref{eq:scalar_power_loss:p_dipole}), and $c_3$ (Eq.~\eqref{eq:scalar_power_loss:c3_dipole}) also contribute at 0PN order as does the scalar quadrupole radiation (Eq.~\eqref{eq:scalar_power_loss:LO_quadropole}).

All combinations of worldline coefficients have been combined to dimensionless parameters and are summarized in Table \ref{tab:dimenionless_params} for the reader's convenience.

We may further use the Kepler relation to express the power losses in terms of the PN expansion parameter $x$. In this case, we obtain
\begin{subequations} \label{eq:scalar_power_loss_x}
    \begin{align} 
        \begin{split} \label{eq:scalar_power_loss:LO_dipole_x}
            P_{\Bar{\phi}}^{l=1, \text{LO}} &= \frac{8}{3 G} \delta q^2 x^4 \;,
        \end{split}\\[2ex]
        \begin{split} \label{eq:scalar_power_loss:p_dipole_x}
            P_{\Bar{\phi}}^{l=1, p_1, p_2} &= \frac{256}{3 G} \Xi_p \, x^5\; ,
        \end{split}\\[2ex]
        \begin{split} \label{eq:scalar_power_loss:c3_dipole_x}
            P_{\Bar{\phi}}^{l=1, c_3} &= \frac{M_1 M_2}{3 \pi} \frac{\msbar - x}{\msbar} \Xi_c\, x^3 \;,
        \end{split}\\[2ex]
        \begin{split} \label{eq:scalar_power_loss:NLO_dipole_x} \raisetag{0.8cm}
            P_{\Bar{\phi}}^{l=1, \text{NLO}} &= \frac{16}{9 G} (-3 + \nu) \delta q^2\,x^5 + \frac{8}{15 G} \msbar^2\, g_2\, \delta q\, x^2 \\
            &- \frac{16}{15 G} (5 g_1 + 3 g_2) \delta q\,x^5\;,
        \end{split}\\[2ex]
        \begin{split} \label{eq:scalar_power_loss:LO_quadropole_x}
            P_{\Bar{\phi}}^{l=2} &= \frac{128}{15 G} \Xi_q^2\,x^5\;.
        \end{split}
    \end{align}
\end{subequations}

Note that we used the additional scalings of $\bar{m}_s \propto x^{3/2}$ and $\Xi_c \propto x^2$ when truncating the above results for the power loss. These scalings arise from conditions that scalar radiation exists and that the perturbative approach is valid (see Sec.~\ref{app:validity}).

Figure~\ref{fig:power_loss_plot} shows the different contributions to the power loss relative to the LO power loss in pure GR. The corresponding model parameters are summarized in Table~\ref{tab::power_loss_plot_params}. 
In the left panel, both (induced) scalar charges are non-zero, a situation typical for NS-NS binaries. 
We observe that the modification of the Kepler relation due to the impact of the scalar field on the binding energy is clearly the most significant effect for the chosen values. 
This contribution to the power loss is present during a large part of the inspiral. 
For small values of $x \lesssim 10^{-4}$, which corresponds to the radial scale where $m_s \sim 1/r$, the exponential suppression of hard scalar interactions becomes sizable. 
The effect of the scalar mass also becomes apparent regarding the radiative dynamics of the system. Here, dipole (quadrupole) radiation is shut off when $m_s < \omega$ ($m_s < \omega/2$). 
Therefore, smaller values of $m_s$ would shift this cutoff to lower $x$, leading to a larger impact of the scalar field during earlier stages of the inspiral. 
In particular, the LO dipole contribution then becomes enhanced as $P^{l=1,\mathrm{LO}}_{\phiBar}/P_\mathrm{GR}^\mathrm{LO} \propto 1/x$.

In the right panel, we set $q_2 = p_2 = 0$, i.e., only one constituent carries a scalar charge. 
Such values would typically be realized in NS-BH inspirals. We immediately notice that the power loss contribution from the modified Kepler relation vanishes. This can be seen from Eq.~\eqref{eq:power_loss_GR} as $P_{\Bar{h}} \propto q_1 q_2$. The same is true for scalar radiation emitted via the cubic self-interaction vertex. In addition, we observe no scalar radiation proportional to induced scalar charges since $P_{\phiBar}^{l=1,p_1,p_2} \propto (p_2 q_1 - p_1 q_2)$. Nonetheless, please note that the power loss from LO dipole radiation is comparable to the left panel, although the scalar charge $q_1$ is an order of magnitude smaller. This is due to the fact that this contribution is proportional to $\delta q^2 \propto (M_2 q_1 - M_1 q_2)^2$, i.e., becomes typically larger if the scalar charges deviate from each other. The same considerations apply to the NLO dipole. The scalar quadrupole, on the other hand, scales with $\Xi_q^2 \propto (M_2^2 q_1 + M_1^2 q_2)^2$, and hence is suppressed for $q_2 = 0$.

\section{Gravitational Wave TaylorF2 Phase} \label{sec:waveform}
We are now set to derive the TaylorF2 phase of the GW signal. In the time domain, a GW $h(t)$ is expressed by the waveform
\begin{align}
    h(t) = A(t) \cos \varphi(t)\;, 
\end{align}
where $A$ denotes the amplitude and $\varphi$ is the phase. We derive the scalar-induced modifications to the TaylorF2 approximant~\cite{Buonanno:2009zt} and thus employ the stationary phase approximation (SPA). For the Fourier transform of the above expression we have
\begin{equation}
    \Tilde{h}(f) = \mathcal{A}(f) e^{i \Psi} \; .
\end{equation}
Here, $\Psi$ solves the following differential equations \cite{Buonanno:2009zt}:
\begin{equation}
    \begin{split}
        0 &= \frac{d \Psi}{d f} - 2 \pi t(f)  \; , \\
        0 &=\frac{d t}{d f} + \frac{G \pi M}{3 v^2} \frac{E'(f)}{P(f)}  \; .
    \end{split}
\end{equation}
Above, $v$ denotes the orbital velocity which is related to the GW frequency $f$ via $v = (\pi G M f)^{1/3}$. Also, the prime denotes differentiation with respect to $v$. It is therefore necessary to calculate the fraction $E'/P$ and integrate twice with respect to the frequency. Here, the binding energy $E$ is given by Eq.~\eqref{eq:binding_energy_full_1PN}. The different contributions to the power loss of the binary are found in Eqs.~\eqref{eq:power_loss_GR} and~\eqref{eq:scalar_power_loss}, respectively.
We perform the computation by expanding $E'/P$ in terms of the perturbative scalar quantities, such as the scalar charge $q$ and self-interaction $c_3$. We then only keep the LO terms, which we calculate up to 0PN order.\footnote{This procedure is formally the same as the expansion around the quadrupolar dominated regime, which was presented in \cite{Sennett:2016klh}. There, the authors expand the inverse power loss around the quadropolar GR emission first, before further expanding in the PN parameter.} Note that this requires to use the pure GR power loss also at 1PN order, which is given by~\cite{Blanchet:2013haa}
\begin{equation}
    P_\text{GR} = \frac{32}{5} \nu x^5 \left[ 1 - \left( \frac{1247}{336} + \frac{35}{12} \nu \right) x \right] .
\end{equation}
We thus have
\begin{equation}
    \Psi(q_1,q_2,\dots) = \Psi_\text{GR} + \Psi_\text{ST} \; ,
\end{equation}
where $\Psi_\text{GR}$ denotes the pure GR phase. $\Psi_\text{ST}$ encodes the corrections due to the scalar field at leading order in the scalar charge, induced charge and self-interaction.
Note that this means that the following results can be added to any existing TaylorF2 template. We further decompose $\Psi_\text{ST}$ to be able to easily identify the components that arise due to the modification to the binding energy \eqref{eq:binding_energy_full_1PN}, which is always present during the inspiral, and the additional scalar radiation, which is only present for sufficient large frequencies. Thus, we decompose
\begin{align}
    \Psi_\text{ST} = \Psi_E &+ \theta \left( \frac{v^3}{\bar{m}_s} - 1 \right) \Psi_{l=1}  \\
    &+ \theta \left( \frac{v^3}{\bar{m}_s} - \frac{1}{2} \right) \Psi_{l=2} \; ,\nonumber
\end{align}
where $\theta$ is the Heaviside step function. $\Psi_E$ denotes the impact of the scalar field onto the gravitational energy loss via the modified Kepler relation. 
The contribution due to direct emission of scalar dipole and quadrupole radiation is captured by $\Psi_{l=1,2}$, respectively. 

We are now ready to present our final results.
The lowest contribution due to radiation arises at $-1$PN order, corresponding to scalar dipole radiation. $\Psi_E$, on the other hand, only contributes from 0PN order. These two corrections read
\begin{widetext}
\begin{subequations} \label{eq:phase_LO}
    \begin{align}
    \begin{split}
        \Psi_E^{\text{0PN}} &= \frac{5 \bar{Q}}{6 \nu  \bar{m}_s^{5/2}} \left[\Gamma \left(\frac{5}{2},\frac{\bar{m}_s}{v^2}\right)+\Gamma \left(\frac{7}{2},\frac{\bar{m}_s}{v^2}\right)+2 \, \Gamma
        \left(\frac{9}{2},\frac{\bar{m}_s}{v^2}\right)\right] \\ 
        &\hspace{5cm} - 
         v^3 \frac{5 \bar{Q}}{6 \nu
         \bar{m}_s^4} \bigg[\Gamma \bigg(4,\frac{\bar{m}_s}{v^2}\bigg)+\Gamma
         \bigg(5,\frac{\bar{m}_s}{v^2}\bigg) + 2\,\Gamma \bigg(6,\frac{\bar{m}_s}{v^2}\bigg)\bigg] \; , 
    \end{split}\\[0.6cm]
        \Psi_{l=1}^{-1\text{PN}} &= \delta  q^2 \frac{5}{896 \nu ^3} \left[-\frac{1}{v^7} \,{}_3F_2\left(-\frac{3}{2},\frac{5}{3},\frac{7}{6} ; \frac{8}{3},\frac{13}{6};\frac{\bar{m}_s^2}{v^6}\right) - \frac{90 \cdot 2^{1/3} 3^{1/2} v^3 \Gamma \left(\frac{2}{3}\right)^3 }{247 \pi \bar{m}_s^{10/3}} + \frac{63 \, \Gamma \left(\frac{1}{3}\right)^3}{256 \cdot 2^{1/3} \pi \bar{m}_s^{7/3}} \right] \; .
    \end{align}
\end{subequations}
Here we utilized the \textit{upper} incomplete gamma function $\Gamma(a, x) = \int_x^\infty t^{a-1} e^{-t} dt$ and ${}_3F_2$ denotes a generalized hypergeometric function.
Note that the above expressions only contain modifications due to the scalar charges $q_1$ and $q_2$.
Corrections due to the induced charges $p_1$ and $p_2$, as well as the self-interaction parameter $c_3$, arise at NLO ($0$PN) with respect to the scalar radiation. Explicitly, these terms are given by
\begin{subequations} \label{eq:phase_NLO}
    \begin{align}
        \begin{split}
            \Psi_{l=1}^{\text{0PN},q} &= -\frac{5 g_2 \delta q \bar{m}_s^2}{9856 \nu^3 v^{11}} \,{}_3F_2\left(-\frac{3}{2},\frac{7}{3},\frac{11}{6} ; \frac{10}{3},\frac{17}{6} ; \frac{\bar{m}_s^2}{v^6}\right) + \frac{45 \Gamma\left(\frac{2}{3}\right)^3 \delta q \left(-1680 g_1- 966 g_2+5 (1064 \nu +659) \delta q\right)}{1605632\ 2^{2/3} \pi  \nu ^3 \bar{m}_s^{5/3}} \\
            &{}\hspace{0.4cm} + \frac{\delta q \left(1680 g_1 + 1008 g_2 - 5 (1064 \nu +659) \delta q\right)}{86016 \nu ^3 v^5} \,{}_3F_2 \left(-\frac{3}{2},\frac{4}{3},\frac{5}{6} ; \frac{7}{3},\frac{11}{6} ; \frac{\bar{m}_s^2}{v^6}\right) \\
            &{}\hspace{0.4cm}+ \frac{5 \sqrt{3} v^3 \Gamma \left(\frac{1}{3}\right)^3 \delta q \left(7728 g_1+ 4368 g_2-23 (1064 \nu + 659) \delta q\right)}{30829568 \sqrt[3]{2} \pi  \nu ^3 \bar{m}_s^{8/3}} \; ,
        \end{split}\\[0.6cm]
        \begin{split}
            \Psi_{l=1}^{\text{0PN},p} &= \Xi_p \frac{5}{16 \nu^3} \left[- \frac{1}{v^5} \,{}_3F_2\left(-\frac{3}{2},\frac{4}{3},\frac{5}{6} ; \frac{7}{3},\frac{11}{6};\frac{\bar{m}_s^2}{v^6}\right) - \frac{6 \cdot 2^{2/3} 3^{1/2} v^3 \Gamma \left(\frac{1}{3}\right)^3 }{187 \pi \bar{m}_s^{8/3}} + \frac{135 \, \Gamma \left(\frac{2}{3}\right)^3}{56 \cdot 2^{2/3} \pi \bar{m}_s^{5/3}} \right] \; ,
        \end{split}\\[0.6cm]
        \begin{split} \label{c3::waveform}
            \Psi_{l=1}^{\text{0PN},c_3} &= \Xi_c \frac{5 G M_1 M_2}{7168 \pi \msbar \nu^3} \left[\frac{1}{v^7} \,{}_3F_2\left(-\frac{3}{2},\frac{5}{3},\frac{7}{6} ; \frac{8}{3},\frac{13}{6};\frac{\bar{m}_s^2}{v^6}\right) + \frac{90 \cdot 2^{1/3} 3^{1/2} v^3 \Gamma \left(\frac{2}{3}\right)^3 }{247 \pi \bar{m}_s^{10/3}} - \frac{63 \, \Gamma \left(\frac{1}{3}\right)^3}{256 \cdot 2^{1/3} \pi \bar{m}_s^{7/3}} \right] \\
            &{}\hspace{0.4cm}+ \Xi_c \frac{25 G M_1 M_2}{55296 \pi \nu^3} \left[ - \frac{1}{v^9} \,{}_3F_2\left(-\frac{3}{2}, 2,\frac{3}{2} ; 3,\frac{5}{2};\frac{\bar{m}_s^2}{v^6}\right) - \frac{24 v^3}{35 \bar{m}_s^4} + \frac{3 \pi}{8 \bar{m}_s^3} \right] \; ,
        \end{split}\\[0.6cm]
        \begin{split}
            \Psi_{l=2}^\text{0PN} &= \Xi_q^2 \frac{1}{32 \nu ^3} \left[-\frac{1}{v^5} \,{}_3F_2\left(-\frac{5}{2},\frac{4}{3},\frac{5}{6} ; \frac{7}{3},\frac{11}{6};\frac{\bar{m}_s^2}{4 v^6}\right) - \frac{720 \cdot 2^{1/3} 3^{1/2} v^3 \Gamma \left(\frac{1}{3}\right)^3 }{4301 \pi \bar{m}_s^{8/3}} + \frac{405 \, \Gamma \left(\frac{2}{3}\right)^3}{112 \pi \bar{m}_s^{5/3}} \right] \; .
        \end{split}
    \end{align}
\end{subequations}
\end{widetext}

Above, we denote contributions arising from a certain scalar field parameter with a respective superscript. For example, $\Psi_{l=1}^{\text{0PN},c_3}$ denotes the contribution of $c_3$ to the $0$PN order phase shift in dipole radiation. For the definition of all rescaled parameters, we refer to Table~\ref{tab:dimenionless_params}.

Note that all hypergeometric functions appear in the form of ${}_3F_2\left(a, b, c\,; b + 1, c + 1\,; z\right)$, with $a,b,c \in \mathbb{Q}$. This can formally be expanded as
\begin{align}
    \begin{split}
        &{}_3F_2\left(a, b, c\,; b + 1, c + 1\,; z\right) = \\
        &\frac{c}{c - b} \,{}_2F_1\left(a, b\,; b + 1 \,; z\right) - \frac{b}{c - b} \,{}_2F_1\left(a, c\,; c + 1\,; z\right).
    \end{split}
    \raisetag{1.3cm}
\end{align}
Since not all math libraries support higher-order hypergeometric functions, this relation might be particularly useful for the implementation of our results.
 
Further, keep in mind that in the above expressions, we use $\hbar = c = 1$. If one wishes to restore these factors in the phase, one may use $\msbar = G M m_s / (\hbar c)$ and replace the combination $G M_1 M_2$ appearing in Eq.~\eqref{c3::waveform} with $G M_1 M_2 / (\hbar c)$. Additionally, it is necessary to substitute all occurrences of the velocity $v$ with $v / c$. 

Let us also note that we have cross-checked all of our computations, in the limit of $m_s = 0$ and $c_3 = 0$, with previous results in massless BD theory \cite{Bernard:2022noq, Sennett:2016klh} using Eq.~\eqref{BDtoSTdictionary}. We find a perfect agreement of the Kepler relation, binding energy, power loss, and the GW phase at the here considered order, i.e., NLO in the PN expansion and LO in the scalar parameters. In addition, we have checked that our expression of the GW phase agrees with the phase computed in \cite{Zhang:2021mks}, corresponding to $m_s \neq 0$, $c_3 = 0$ at LO in the PN expansion.

The corrections due to the additional scalar degree of freedom can now readily be combined with pure GR TaylorF2 templates at arbitrary precision \cite{Buonanno:2009zt}. Further, since we kept the scalar parameters generic up to this point, our results can easily be adapted to a large class of scalar-tensor theories. All there is left to do is adjust the potential parameters and worldline couplings to a desired model. To facilitate using our results, we provide a link to a repository in the supplemental material \cite{waveform_code_snippet} which contains a Python implementation of Eqs.~\eqref{eq:phase_LO}~and~\eqref{eq:phase_NLO}.

\section{Conclusions \label{sec:conclusions}}
In this work, we have for the first time derived the NLO gravitational TaylorF2 phase of inspiraling binary systems in GR augmented by a massive, self-interacting scalar field. 
To this end, we have employed the scale hierarchy during the inspiral to treat the binary within an EFT framework. 
We have calculated the conservative and dissipative dynamics of the system at 1PN order to ultimately determine the modifications of the gravitational phase up to $0$PN order. This corresponds to NLO in the scalar radiation.  
This result can now readily be combined with pure GR templates---in the TaylorF2 approximation---at arbitrary precision.

Throughout our work, we have chosen a largely model-independent ansatz by choosing a generic, polynomial scalar potential and leaving the scalar-wordline couplings undetermined. 
This allows us to match our result to a multitude of new physics scenarios by adjusting the Taylor coefficients, including extensions of GR as well as of the SM of particle physics, provided an appropriate coupling to the binary worldlines. 

Ultimately, the goal is to set constraints on new physics using data from GW observatories. 
Therefore, we will as a next step use data from GW170817 to test various scalar-tensor theories. 
Once new data is available in the future, we will then be able to easily repeat the procedure, further narrowing down the window of new physics. 

It will also be interesting to extend the calculations to elliptic orbits, thus allowing us to utilize, e.g.,~pulsar data.

From the EFT point of view, we will extend our computations to take into account scalar effects which enter at a higher order in the velocity expansion, further pushing the precision of our predictions. 
In addition, our computations may be further generalized by, e.g.,~including vector degrees of freedom. These challenging tasks are left for future work.

\begin{acknowledgments}
The authors thank N.~Becker, E.~Genoud-Prachex, S.~Ghosh, A.~Kuntz, S.~Pal, Y.~Schaper, S.~Tsujikawa, and J.~Zhang for useful discussions, and D.~Trestini for helpful comments on the first version of this manuscript. RFD and DS acknowledge support by the Deutsche Forschungsgemeinschaft (DFG, German Research Foundation) through
the CRC-TR 211 `Strong-interaction matter under extreme conditions' -- project number 315477589 -- TRR 211. LS acknowledges the support by the State of Hesse within the Research Cluster ELEMENTS (Project ID 500/10.006).
\end{acknowledgments}

\onecolumngrid
\appendix

\cftlocalchange{toc}{309pt}{0cm}
\cftaddtitleline{toc}{section}{\textbf{Appendices}}{}
\cftlocalchange{toc}{1.55em}{2.55em}

\section{Redundant Operators} \label{app:redundant_operators}
In general, there are many additional worldline couplings of the scalar field beyond the ones shown in Eq.~\eqref{eq:Spp}. In fact, Appendix A of \cite{Damour:1998jk} lists these couplings explicitly up to terms containing second derivatives in the scalar field $\phi$ and the metric $g_{\mu\nu}$. Here, it is important to note that at this point the authors did not employ any power counting rules, i.e., these operators do not necessarily contribute at 1PN order. They arrive at the general wordline action
\begin{align}
    \begin{split}
        S_{pp} = - \int d \tau \bigg[ m(\phi) &+ I(\phi) R + J(\phi) u^\mu u^\nu R_{\mu\nu} + K(\phi) \square \phi + L(\phi) u^\mu u^\nu \nabla_\mu \partial_\nu \phi \\
        &+ M(\phi) u^\mu u^\nu \partial_\mu \phi \partial_\nu \phi + N(\phi) g^{\mu\nu} \partial_\mu \phi \partial_\nu \phi \bigg] \; .
    \end{split}
\end{align}
After a series of redefinitions and neglecting operators which cannot contribute at orders lower than $v^4$, one obtains an equally valid action\begin{equation}
    S_{pp} = - \int d \tau \bigg[ m(\phi) + \Tilde{N}(\phi) g^{\mu\nu} \partial_\mu \phi \partial_\nu \phi \bigg] \; ,
\end{equation}
where $\Tilde{N} = N + \alpha(\phi) L + 2 I$ and $\alpha(\phi) = \partial \ln m(\phi) / \partial \phi$. It is thus directly shown that all other operators are either redundant or do not contribute at 1PN order, and hence do not have to be considered. Also, the second term can be identified as the LO finite-size effect due to an external tidal field. In order to further investigate at which PN order these operators contribute, it is useful to expand $m(\phi)$ and $\Tilde{N}(\phi)$ in a Taylor series:
\begin{equation}
    m(\phi) = m(0) + q \frac{\phi}{\MPl} + p \frac{\phi^2}{\MPl^2} + \mathcal{O}(\phi^3)\;, \quad \text{and} \quad \Tilde{N}(\phi) = \Tilde{N}(0) + \Tilde{N}'(0) \frac{\phi}{\MPl} + \mathcal{O}(\phi^2)\;.
\end{equation}
Here, primes denote the derivative with respect to the scalar field, and we have defined $q = \MPl\, m'(0)$ and $p = \MPl^2\, m''(0)$. It is straightforward to identify $m(0)$ with the usual GR mass, while the higher-order derivatives define the couplings of the scalar field to the worldline. Employing the power counting rules, it is now easy to see that at 1PN order the expansion of $m(\phi)$ can be truncated beyond the quadratic term. Likewise, it can be shown that the expansion of $N(\phi)$ contributes only from 3PN order for compact objects. The impact of the second term on the waveform has been investigated in Ref.~\cite{Bernard:2019yfz} for massless scalar fields.

Similarly, by expanding the self-interaction potential of the scalar field in a power series, it can be shown that at 1PN order only the cubic self-interaction is relevant. In fact, a self-interaction term of the form $\phi^n$ becomes relevant at $(n-2)$PN order in the expansion.

\section{Validity of the EFT} \label{app:validity}
Let us comment on the assumptions we make in order for the EFT prescription to be valid. Firstly, throughout the calculations we assumed that at LO the inspiral is described by Newtonian gravity.
This assumption is, e.g., extensively used when establishing the power counting rules. 
To justify this perturbative treatment of the binary system, the Newtonian potential has to remain the LO term even when adding the scalar field; i.e., the scalar exchange has to be suppressed. This directly gives a constraint on the scalar charges
\begin{equation}
    \frac{|q_n|}{M_n} < 1 \;, \quad \text{and likewise} \quad \frac{|p_n|}{M_n} < 1 \; .
\end{equation}

Further, for the cubic self-coupling $c_3$ we note that the three-scalar vertex scales $\sim c_3 \MPl^2 r^2 L^{-1/2} v^2$. Higher-order terms in $c_3$ scale with higher powers of $c_3 \MPl^2 r^2$. We thus also demand that $|c_3| \MPl^2 r^2 < 1$, which implies that 
\begin{equation}
    |c_3| \frac{M^2}{\MPl^2} < \gamma^2 < 1 \; .
\end{equation}
With $M \sim M_\odot$ this implies $|c_3| < 10^{-80}$, which was also noted earlier \cite{Porto:2007pw}. However, it is important to keep in mind that the above constraint only applies for the case when the scalar field behaves effectively massless (i.e., $m_s r \ll 1$), meaning that it is to be understood as a conservative upper bound on $|c_3|$. If $m_s r \ll 1$ is never achieved during the inspiral phase, the constraint could be relaxed, since additional exponential suppression in $m_s r$ helps to subdue higher-order $c_3$ terms. In the final expressions for the Lagrangian, binding energy, power loss, and TaylorF2 phase, we always have a combination of the form $c_3 M^2 / \MPl^2$. Therefore, if $c_3$ fulfills the above-mentioned bound, its effect is still relevant for the dynamics. While Ref.~\cite{Porto:2007pw} considered this constraint to be resulting in the need for unnatural fine-tuning on $c_3$, it is indeed naturally fulfilled in several modified gravity models, such as $R^2$ gravity (see e.g., \cite{Capozziello:2011et, DeFelice:2010aj}), where the Einstein-Hilbert action is supplemented with a term quadratic in the Ricci scalar. After performing a transformation to the Einstein frame one can isolate an additional scalar degree of freedom with a potential of the form \cite{DeFelice:2010aj, Staykov:2014mwa}
\begin{equation}
    V(\phi) \sim m_s^2 \phi^2 - \frac{m_s^2}{\MPl} \phi^3 + \mathcal{O}(\phi^4) \; ,
\end{equation}
where we have dropped $\mathcal{O}(1)$ pre-factors for brevity.
Hence $|c_3| \sim m_s^2 / \MPl^2$, which is much less than $10^{-80}$ for $m_s < 10^{-13}$\,eV, thus fulfilling the above constraint. If $m_s > 10^{-13}$\,eV, then $m_s r < 1$ is never achieved during the inspiral and hence all contributions of $c_3$ are highly suppressed, meaning that in either case one expects a perturbative treatment to be valid.

When applying the here obtained results for, e.g.,~the power loss, we also implicitly assume that the effects from higher-order worldline couplings and higher-order self-interactions in the scalar field are suppressed. It is reasonable to presume, on grounds of the established power counting rules, that those terms would only contribute to higher-order PN orders. However, the numerical values of the parameter accompanying those higher-order operators might be large enough to still cause relevant effects even at lower orders in the PN expansion. For example, this is the case with the tidal deformability in pure GR, which formally contributes at 5PN order for non-spinning objects, but might be relevant even for lower-order waveforms, if its numerical value is sufficiently large.

Lastly, we assume that the worldline coupling coefficients are constant during the inspiral. However, it is expected that at higher order in the expansion it is necessary to consider time-dependent couplings \cite{Goldberger:2012kf}.

\section{Renormalization} \label{app:renormalization}
In this section, we discuss some technical aspects regarding the renormalization procedure. In the computation of the binding energy (cf.~Sec.~\ref{sec:scalar_tensor_eft}) we have neglected pure self-force diagrams, shown in Fig.~\ref{renormalization::1PN}. Here we show that these diagrams solely contribute to the renormalization of the worldline couplings, thus can be absorbed by a redefinition of the gravitational masses $M_i$ and scalar charges $q_i$. 

Since we employ dimensional regularization, diagrams such as the right one in Fig.~\ref{renormalization::massrenorm} simply vanish, since they contain scale-less and power-law divergent integrals, i.e.,
\begin{align}
\includegraphics[scale=0.85,valign=c]{Renorm_figure1.pdf}
        \; \propto \int \frac{d^3 \kbf}{(2\pi)^3} \frac{1}{\kbf^2} \xrightarrow{\rm dim.~reg.} 0 \; .
\end{align}
However, couplings of the worldline to itself via a scalar propagator do not vanish identically in this scheme since they contain the mass of the scalar particle as a physical scale. As such, the left diagram in Fig.~\ref{renormalization::massrenorm} evaluates to 
\begin{equation}
    \begin{split}
\includegraphics[scale=0.85,valign=c]{Renorm_figure0.pdf} &= \frac{1}{2} \left( \frac{-i q}{\MPl} \right)^2 \int dt \int \frac{d^3\kbf}{(2 \pi)^3} \frac{-i}{\kbf^2 + m_s^2} \\
     & = \frac{i}{2} \left( \frac{-i q}{\MPl} \right)^2 \int dt \frac{m_s}{4 \pi} = - i \underbrace{\frac{m_s\, q^2}{8 \pi \MPl^2}}_{\equiv \delta M_\text{sf}} \int dt \; .
    \end{split}
\end{equation}
Note the factor of $1/2$ in front, which arises from the expansion of the action and keeping two identical vertices. Also, we here used
\begin{equation}
    \int \frac{d^3 \kbf}{(2\pi)^3} \frac{1}{\kbf^2 + m_s^2} \xrightarrow{\rm dim.~reg.} - \frac{m_s}{4 \pi} \;.
\end{equation}
Thus, the above diagram effectively shifts the mass of the neutron star by $\delta M_\text{sf}$. We denote the mass $M$ appearing in $\Spp$ as the \textit{bare} mass $M_b$. Likewise, in the following we denote all other bare parameters, i.e.,~those parameters that directly appear in the expansion of the point-particle action, with a subscript `b'. Then, at lowest order, the bare mass is shifted to the \textit{physical} mass with
\begin{equation}
\label{renormalization::physicalMass}
    M_{\rm phys} = M_{\rm b} + \delta M_\text{sf} = M_{\rm b} + \frac{m_s\, q_\text{b}^2}{8 \pi \MPl^2} \; .
\end{equation}
By calculating all diagrams in which a single graviton is coupled to the second NS we should get the same result for the renormalized mass of the first neutron star as the one given in Eq.~\eqref{renormalization::physicalMass}. 
Contributing diagrams are the ones given on the left in Fig.~\ref{renormalization::singlegravitonvertex} and 
in Fig.~\ref{fig:binding_energy:scalar}\,v), respectively. Indeed, the first one evaluates to $\sim 2 \delta M_\text{sf}$ and the second contributes with $\sim -\delta M_\text{sf}$, thus yielding the same total mass shift as in Eq.~\eqref{renormalization::physicalMass}.

In addition to the mass, the scalar charge also obtains corrections due to self-force diagrams shown in Fig.~\ref{renormalization::singlescalarvertex}. Evaluating this integral leads to 
\begin{equation}
    \delta q_{\rm sf} = m_s \frac{q_\text{b}\,p_\text{b}}{4 \pi \MPl^2} \;.
\end{equation}
However, additional contributions are coming from diagrams in Figs.~\ref{fig:binding_energy:scalar}\,vi) and \ref{fig:binding_energy:scalar}\,vii), which are not pure self-force diagrams (in the sense that they do not only contain diverging integrals). In total, we obtain
\begin{align}
\label{renormalization::phyiscalCharge}
    \begin{split}
        q_{\rm phys} &= q_\text{b} + \delta q_{\rm sf} + \frac{ q_\text{b}^2 c_3 \log 3}{16 \pi m_s} - \frac{m_s M_\text{b} q_\text{b} \,\gamma}{32 \pi \MPl^2} \\
        &= q_\text{b} + \frac{q_\text{b}}{4\pi} \bigg[ \frac{p_\text{b}\,m_s}{\MPl^2} + \frac{q_\text{b} \, c_3 \log 3}{4 \, m_s} - \frac{m_s M_\text{b}\,\gamma}{8 \MPl^2}\bigg] \;,
    \end{split}
\end{align}
with $\gamma$ denoting the Euler-Mascheroni constant and $\log$ denoting the natural logarithm.

In summary, by expressing the \textit{bare} charge and mass in terms of the \textit{physical} charge and mass via Eqs.~\eqref{renormalization::physicalMass} and \eqref{renormalization::phyiscalCharge}, we automatically cancel all pure self-force diagrams appearing in Fig.~\ref{renormalization::1PN} and additionally also all redundant terms that appear when evaluating the diagrams in Fig.~\ref{fig:binding_energy_diagrams}.

\section{Radiated Energy in Multipole Moments of the Scalar Field}
\label{app:energy_loss_scalar}

\begin{figure}[h]
    \centering
    \includegraphics[scale=1.5]{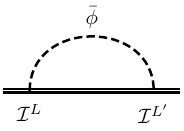}
    \caption{Leading order linear emission diagram, which allows calculation of the power emitted in scalar radiation. Here $\mathcal{I}^L$ denotes the $L$ multipole moment of the linear source term.}
    \label{app:radiation_moments_scalar}
\end{figure}
Let us outline how to compute the power loss via scalar radiation. The general procedure to perform the multipole expansion of a massless scalar field without self-interactions is described in Ref.~\cite{Ross:2012fc}. Here, a general formula is presented to obtain the emitted power in arbitrary moments due to the LO linear emission diagram in Fig.~\ref{app:radiation_moments_scalar}. Since we here consider a massive, self-interacting scalar field, we appropriately modify the derivations of \cite{Ross:2012fc} to obtain adapted expressions for the power loss. For the convenience of the reader, we also repeat some of the initial steps of the calculation presented in \cite{Ross:2012fc}, but refer to this publication for the full amount of details.

Considering a linear source term, the relevant part of the scalar field action can be written as
\begin{equation}
    S = \int dx^4 \left( \frac{1}{2} \partial_\mu \phi \partial^\mu \phi + \frac{1}{\MPl} J \phi \right) \; .
\end{equation}
We assume the source to be localized and Taylor expand the field coupled to $J$ at the origin, which we choose to be inside the source:
\begin{equation}
    \phi(t, \xbf) = \sum_{n=0}^\infty \frac{\xbf^N}{n!} (\partial_N \phi) (t, 0) \; .
\end{equation}
Inserting this expansion into the linear source term and decomposing the source moments into symmetric and trace free (STF) tensors, we further obtain
\begin{equation} \label{eq::source_term}
    S_\text{src} =  \frac{1}{\MPl} \int dx^4 J \phi =  \frac{1}{\MPl} \int dt \sum_{l = 0}^\infty \frac{1}{l!} \sum_{p = 0}^\infty \frac{(2l + 1)!!}{(2p)!! (2l + 2p + 1)!!} \int d^3 \xbf J r^{2p} \xbf_\text{STF}^L (\nabla^2)^p \partial_L \phi \; .
\end{equation}
At this point Ref.~\cite{Ross:2012fc} utilizes that outside the source $\square \phi \equiv \partial_t^2 \phi - \nabla^2 \phi = 0$, and thus $\nabla^2 \phi = \partial_t^2 \phi$, in order to turn the contracted spatial derivatives acting on the last term into time derivatives. However, as was also recently noted in Ref.~\cite{Bhattacharyya:2023kbh}, in the case of non-vanishing potential the equation of motion instead reads $\square \phi \equiv \partial_t^2 \phi - \nabla^2 \phi = - V'(\phi)$, and hence $\nabla^2 \phi = \partial_t^2 \phi + V'(\phi)$. Since we only consider linear emission diagrams, we can truncate all non-linear terms in $V'$; i.e., only the mass term is relevant. Therefore, Eq.~\eqref{eq::source_term} becomes
\begin{equation}
    S_\text{src} =  \frac{1}{\MPl} \int dt \sum_{l = 0}^\infty \frac{1}{l!} \sum_{p = 0}^\infty \frac{(2l + 1)!!}{(2p)!! (2l + 2p + 1)!!} \int d^3 \xbf J r^{2p} \xbf_\text{STF}^L (\partial_t^2 + m_s^2)^p \partial_L \phi \; .
\end{equation}
Partial integration with respect to time now also allows having time derivatives acting on the source term $J$ instead of the field $\phi$. We can thus write 
\begin{equation}
    S_\text{src} = \frac{1}{\MPl} \int dt \sum_{l = 0}^\infty \frac{1}{l!} \mathcal{I}^L \partial_L \phi \; ,
\end{equation}
with the multipole moments $\mathcal{I}^L$ given by
\begin{equation}
    \mathcal{I}^L = \sum_{p = 0}^\infty \frac{(2l + 1)!!}{(2p)!! (2l + 2p + 1)!!} \int d^3 \xbf\,r^{2p} \xbf_\text{STF}^L \,(\partial_t^2 + m_s^2)^{p} J \; .
\end{equation}
Up to 0PN order, the relevant terms are given by
\begin{equation}
    \mathcal{I}^i_{p=0} = \int d^3 \xbf\, \xbf^i J \; , \quad \mathcal{I}^i_{p=1} = \frac{1}{10} \int d^3 \xbf\, \xbf^i (\partial_t^2 + m_s^2) r^2 J \; , \quad \text{and} \quad \mathcal{I}^{ij}_{p=0} = \int d^3 \xbf \left[\xbf^i \xbf^j \right]_\text{STF} J \; .
\end{equation}
We now turn to the calculation of the emitted power. The diagram shown in Fig.~\ref{app:radiation_moments_scalar} evaluates to
\begin{equation}
\begin{split}
    \includegraphics[scale=0.8,valign=c]{SF_figure0.pdf} \hspace{0.5cm} &=  \frac{1}{\MPl^2} \frac{1}{l!\,\Tilde{l}!} \int dt_1 \int dt_2\, \mathcal{I}^L(t_1) \mathcal{I}^{\Tilde{L}}(t_2) \int \frac{d^4k}{(2\pi)^4} \frac{e^{i k_0 (t_1 - t_2)}}{k^2 - m^2 + i \epsilon} \kbf^L \kbf^{\Tilde{L}} \\
    &= \frac{1}{\MPl^2} \frac{1}{l!\,\Tilde{l}!} \int \frac{d^4k}{(2\pi)^4}  \frac{1}{k^2 - m^2 + i \epsilon} \mathcal{I}^L(k_0) \mathcal{I}^{\Tilde{L}}(k_0)^* \kbf^L \kbf^{\Tilde{L}} \\
    &= \frac{1}{\MPl^2} \frac{1}{l!\,\Tilde{l}!} \int \frac{k_0}{2\pi} \int d \Omega \int \frac{d k}{(2\pi)^3} k^{2 + l + \Tilde{l}} \nbf^L \nbf^{\Tilde{L}}  \frac{1}{k_0^2 - k^2 - m^2 + i \epsilon} \mathcal{I}^L(k_0) \mathcal{I}^{\Tilde{L}}(k_0)^*
    \; ,
\end{split}
\end{equation}
where $\nbf = \kbf / |\kbf|$. In the second line, we have expressed the moments $\mathcal{I}$ in terms of their Fourier modes, and subsequently switched to spherical coordinates. The surface integral results in
\begin{equation}
    \int d \Omega\,\nbf^L \nbf^{\Tilde{L}} = \delta^{l \Tilde{l}} \frac{4 \pi}{(2l + 1)!!} l! \; ,
\end{equation}
such that
\begin{equation}
\begin{split}
    \includegraphics[scale=0.8,valign=c]{SF_figure0.pdf} \hspace{0.5cm} 
    &= \frac{1}{\MPl^2} \frac{1}{l! (2l + 1)!!} \frac{1}{4 \pi^3} \int dk_0 \int dk\, k^{2l + 2} \left| \mathcal{I}^L(k_0) \right|^2 \frac{1}{k_0^2 - k^2 - m^2 + i \epsilon} \\
    &= \frac{1}{\MPl^2} \frac{1}{l! (2l + 1)!!} \frac{i}{4 \pi^2} \int dk_0 (k_0 - m^2)^{l + 1/2} \left| \mathcal{I}^L(k_0) \right|^2 \; .
    \end{split}
\end{equation}
In the second line, we have solved the $dk$ integral by closing the contour around the pole $k = + \sqrt{k_0^2 - m^2}$ in the upper plane.
The emitted energy is thus given by
\begin{equation}
    P = \frac{1}{l! (2l + 1)!!} \frac{1}{4 \pi^2 \MPl^2 T} \int d \omega \left(1 - \frac{m_s^2}{\omega^2}\right)^{l + 1/2} \omega^{2l + 2} \left| \mathcal{I}^L(\omega) \right|^2,
\end{equation}
where $T = 2 \pi \delta(0)$ and $k_0 = \omega$. The same result was presented in Ref.~\cite{Huang:2018pbu}, but without elaborating on the calculation and the authors did not mention that the multipole moments also have to be modified in the massive case.

\section{Computation of the Feynman Diagrams} \label{app:feynman_diagrams}
In this section, we explicitly compute the Feynman diagrams which are shown in the main text. First, let us give a general recipe on how to compute the diagrams to a certain PN order with the power counting scheme from Sec.~\ref{sec:EFT}. A similar discussion is found in Ref.~\cite{Kuntz:2019zef}.

\begin{enumerate}
    \item 
    Expand the full action of the theory \eqref{eq:action} in powers of the velocity $\vbf$, as shown in Sec.~\ref{sec:scalar_tensor_eft}. This renders all vertices between the NS worldlines, the scalar field, as well as the graviton field.
    \item
    Decompose the fields into radiation and potential modes. From the power counting rules derived in Sec.~\ref{sec:EFT}, identify the PN order at which a vertex contributes.
    \item 
    From the action expansion, read off the Feynman rules for the given interactions. These are multiplied by a factor $i$ from the expansion of the path integral. An overview of the relevant interactions at 1PN order, together with their corresponding velocity scalings and Feynman rules, is given in Tables~\ref{feynmanRules::bindingEnergy}~and~\ref{feynmanRules::radiation}.
    \item
    Draw all diagrams that contribute to the desired PN order. Here, consider only diagrams which remain connected when the wordlines are removed. Note that potential modes can only appear as internal lines, while radiation modes solely enter as external legs. For any internal line, multiply by the respective propagator. 
    In addition, neglect quantum loop diagrams as they are suppressed by $\hbar/L$~\cite{Goldberger:2004jt}, with $L$ the orbital angular momentum of the system.
    \item
    Collect the combinatorial factor for each diagram. This includes both pre-factors from the expansion of the path integral, as well as symmetry factors from the Wick contractions. 
\end{enumerate}\vfill

\subsection{Feynman Rules}\label{app:feynman_rules}

\bgroup
\setlength\cellspacetoplimit{5pt}
\setlength\cellspacebottomlimit{5pt}
\setlength\tabcolsep{5pt}
\begin{table}[H]
\centering
\begin{tabular}{ | Sc | C{2cm} | C{11cm} | }
    \hline
    \bf{Diagram} & \bf{Scaling} & \bf{Expression} \\
    \hline
    {\cincludegraphics{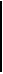}}
    &
    $\displaystyle \sim L$
    & 
    $\displaystyle i \int dt \frac{M}{2} \vbf^2$ \\ \hline     
    {\cincludegraphics{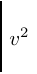}}
    & $\displaystyle \sim L v^2$ &  $ \displaystyle i \int dt \frac{M}{8} \vbf^4$ \\ \hline \hline
    {\cincludegraphics{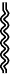}}
    & $\displaystyle \sim 1$ &  $ \displaystyle -(2\pi)^3 \frac{i}{|\kbf|^2} \delta^{(3)}(\kbf + \qbf) \delta(t_1 - t_2) P_{\mu \nu; \alpha \beta}$ \\ \hline
    
    {\cincludegraphics{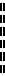}} 
    & $\displaystyle \sim 1$ &  $ \displaystyle -(2\pi)^3 \frac{i}{|\kbf|^2 + m^2} \delta^{(3)}(\kbf + \qbf) \delta(t_1 - t_2) $ \\ \hline 
    
    {\cincludegraphics{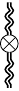}} 
    & $\displaystyle \sim v^2$ &  $ \displaystyle -(2\pi)^3 \frac{i}{|\kbf|^4} \delta^{(3)}(\kbf + \qbf) \frac{\partial^2}{\partial t_1 \partial t_2}\delta(t_1 - t_2) P_{\mu \nu; \alpha \beta} $ \\ \hline 
    
    {\cincludegraphics{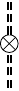}} 
    & $\displaystyle \sim v^2$ &  $ \displaystyle -(2\pi)^3 \frac{i}{(|\kbf|^2 + m^2)^2} \delta^{(3)}(\kbf + \qbf)\frac{\partial^2}{\partial t_1 \partial t_2}\delta(t_1 - t_2) $ \\ \hline 
    \hline
    
    {\cincludegraphics{FRP_figure6.pdf}} & $\displaystyle \sim L^\frac{1}{2}$ &  $ \displaystyle -i \frac{M}{2 \MPl}  \int dt \int_\kbf \exp\left(i \kbf \xbf \right) \eta^{0\mu} \eta^{0\nu}$ \\ \hline 
    
    {\cincludegraphics{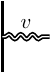}} & $\displaystyle \sim L^\frac{1}{2} v$ &  $ \displaystyle -i \frac{M}{\MPl}  \int dt \int_\kbf  \exp\left(i \kbf \xbf \right) \vbf^i \eta^{0(\mu} \eta^{\nu)}_{\;i}$ \\ \hline
    
    {\cincludegraphics{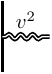}} & $\displaystyle \sim L^\frac{1}{2} v^2$ &  $ \displaystyle -i \frac{M}{2\MPl}  \int dt \int_\kbf  \exp\left(i \kbf \xbf \right) \left(\eta_i^\mu \eta_j^\mu \vbf^i \vbf^j + \frac{1}{2} \eta^{0\mu} \eta^{0\nu} \vbf^2 \right)$ \\ \hline
    
    {\cincludegraphics{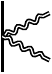}} & $\displaystyle \sim v^2$ &  $ \displaystyle i {\frac{M}{8\MPl^2}} \int dt \int_{\kbf,\qbf⁄}  \exp\left(i (\kbf + \qbf) \xbf \right) \eta^{0\mu} \eta^{0\nu} \eta^{0\lambda} \eta^{0\sigma}$ \\ \hline \hline
    
    {\cincludegraphics{FRP_figure10.pdf}} & $\displaystyle \sim \frac{q}{M} L^\frac{1}{2}$ &  $\displaystyle -i \frac{q}{\MPl} \int dt \int_{\kbf} \exp\left(i \kbf \xbf \right)$ \\ \hline
    
    {\cincludegraphics{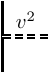}} & $\displaystyle \sim \frac{q}{M} L^\frac{1}{2} v^2$ &  $\displaystyle i \frac{q}{2\MPl} \int dt \int_{\kbf} \exp\left(i \kbf \xbf \right) \vbf^2$ \\ \hline
    
    {\cincludegraphics{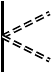}} & $\displaystyle \sim \frac{p}{M} v^2$ &  $\displaystyle -i \frac{p}{\MPl^2} \int dt \int_{\kbf, \qbf} \exp\left(i (\kbf + \qbf) \xbf \right)$ \\ \hline 
\end{tabular}
\end{table}

\begin{table}
\begin{tabular}{ | Sc | C{2cm} | C{11cm} | }
\hline
    {\cincludegraphics{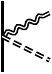}} & $\displaystyle \sim \frac{q}{M} v^2$ &  $\displaystyle -i \frac{q}{2 \MPl^2} \int dt \int_{\kbf, \qbf} \exp\left(i (\kbf + \qbf) \xbf\right)$ \\ \hline \hline
     
    {\cincludegraphics{FRP_figure14.pdf}} & $\displaystyle \sim L^{-\frac{1}{2}} v^2$ &  $\displaystyle -\frac{i}{4 \MPl} \delta(t_1 - t_2) \delta(t_1 - t_3) (2\pi)^3 \delta^{(3)}\left(\sum_r \kbf_r\right) \prod_r \frac{i}{\kbf_r^2} \sum_r \kbf_r^2 $ \\ \hline
    
    {\cincludegraphics{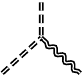}} & $\displaystyle \sim \frac{m_s^2}{\MPl} \frac{r^2  v^2}{L^\frac{1}{2}}$ &  $\displaystyle -i \frac{m_s^2}{4 \MPl} \delta(t_1-t_2) \delta(t_1-t_3) (2\pi)^3 \delta^{(3)}\left(\sum_r \kbf_r\right) \left( \prod_{r=1,2} \frac{i}{\kbf_r^2 + m^2} \right) \frac{i}{\kbf_3^2} \eta^{\mu\nu} P_{00;\mu\nu}$ \\ \hline
    
    {\cincludegraphics{FRP_figure16.pdf}} & $\displaystyle \sim c_3 \frac{\MPl r^2}{L^\frac{1}{2}} v^2 $ &  $\displaystyle -i \frac{c_3}{3!} \delta(t_1-t_2) \delta(t_1-t_3) (2\pi)^3 \delta^{(3)}\left(\sum_r \kbf_r\right) \prod_r \frac{i}{\kbf_r^2 + m^2}$ \\ \hline
  \end{tabular}
  \caption{Feynman rules necessary to compute the binding energy diagrams at 1PN order.}
  \label{feynmanRules::bindingEnergy}
\end{table}

\begin{table}[H]
  \centering
    \begin{tabular}{ | Sc | C{2cm} | C{11cm} | }
    \hline
    \bf{Diagram} & \bf{Scaling} & \bf{Expression} \\
    \hline
    {\cincludegraphics{FRR_figure0.pdf}}
    &
    $\displaystyle \sim L^\frac{1}{2} v^\frac{1}{2}$
    & 
    $\displaystyle -i \frac{M}{2 \MPl} \int dt  \; \Bar{h}_{00}$ \\ \hline     

    {\cincludegraphics{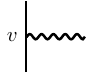}}
    & $\displaystyle \sim L^\frac{1}{2} v^\frac{3}{2}$ &  $\displaystyle -i \frac{M}{\MPl} \int dt \; \Bar{h}_{0i} \vbf^i$ \\ \hline 

    {\cincludegraphics{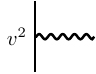}}
    & $\displaystyle \sim L^\frac{1}{2} v^\frac{5}{2}$ &  $\displaystyle -i \frac{M}{2 \MPl} \int dt \left(\bar{h}_{ij} \vbf^i \vbf^j + \frac{\bar{h}_{00}}{2} \vbf^2 \right)$ \\ \hline \hline

    {\cincludegraphics{FRR_figure3.pdf}}
    & $ \displaystyle \sim \frac{q}{M} L^\frac{1}{2} v^\frac{1}{2}$ & $\displaystyle -i \frac{q}{\MPl} \int dt \; \phiBar$ \\ \hline 

    {\cincludegraphics{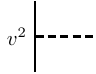}}
    & $\displaystyle \sim \frac{q}{M} L^\frac{1}{2} v^\frac{5}{2}$ &  $\displaystyle -i \frac{q}{2 \MPl} \int dt \; \vbf^2 \phiBar$ \\ \hline \hline 

    {\cincludegraphics{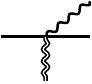}}
    & $\displaystyle \sim v^\frac{5}{2} $ &  $\displaystyle -i \frac{M}{4 \MPl^2} \int dt \int_\kbf \exp\left(i \kbf \xbf\right) \eta^{0\mu} \eta^{0\nu} \Bar{h}_{00}$ \\ \hline 

    {\cincludegraphics{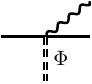}}
    & $\displaystyle \sim \frac{q}{M} v^\frac{5}{2}$ &  $\displaystyle -i \frac{q}{2\MPl^2} \int dt \int_\kbf \exp\left(i \kbf \xbf \right) \bar{h}_{00}$ \\ \hline

    {\cincludegraphics{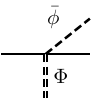}}
    & $\displaystyle \sim \frac{p}{M} v^\frac{5}{2}$ &  $\displaystyle -i \frac{2 p}{\MPl} \int dt \int_\kbf \exp\left(i \kbf \xbf \right) \phiBar$ \\ \hline
\end{tabular}
\end{table}

\begin{table}[H]
\begin{tabular}{ | Sc | C{2cm} | C{11cm} | }
\hline
    {\cincludegraphics{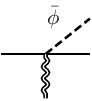}}
    & $\displaystyle \sim \frac{q}{M} v^\frac{5}{2}$ &  $\displaystyle -i \frac{ q}{2 \MPl^2} \int dt \int_\kbf \exp\left(i \kbf \xbf \right) \eta^{0\mu} \eta^{0\nu} \phiBar$ \\ \hline \hline 

    {\cincludegraphics{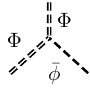}} & $\displaystyle \sim c_3 \frac{M r v^\frac{1}{2}}{L^\frac{1}{2}} $ &  $\displaystyle -i \MPl \frac{3 c_3}{3!} \delta(t-t_1) \delta(t-t_2) (2\pi)^3 \frac{i}{\kbf_1^2 + m^2} \frac{i}{\kbf_2^2 + m^2}  \phiBar$ \\ \hline

    {\cincludegraphics{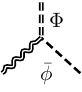}} & $\displaystyle \sim \msbar^2 \frac{1}{L^\frac{1}{2} v^\frac{3}{2}} $ &  $\displaystyle -i \frac{2 m_s^2}{4 \MPl}  \delta(t-t_1) \delta(t-t_2) (2\pi)^3 \frac{i}{\kbf_1^2 + m^2} \frac{i}{\kbf_2^2} \eta^{\mu\nu} P_{00;\mu\nu} \phiBar$ \\ \hline

  \end{tabular}
  \caption{Feynman rules involving the radiation modes, up to 1PN order.}
  \label{feynmanRules::radiation}
\end{table}
\egroup

\subsection{Conservative Dynamics} \label{app:binding_energy_computation}
We start with the calculation of the conservative dynamics. This comprises all diagrams at 1PN which only contain hard scalar and graviton modes (cf.~Fig.~\ref{fig:binding_energy_diagrams}). 
\begingroup
\addtolength{\jot}{.8em}
\begin{align}
    &\begin{aligned}
        \includegraphics[scale=0.8,valign=c]{BE_figure0.pdf}
        &= i \frac{M_1 M_2}{4 \MPl^2} \int dt_1 \int dt_2 \delta(t_1-t_2) \int \frac{d^3\kbf}{(2\pi)^3} \frac{\exp(-i \kbf (\xbf_1 - \xbf_2))}{\kabsbf^2} P_{00;00}  = i \frac{M_1 M_2}{32 \pi \MPl^2} \int dt \frac{1}{r}\\
        &= i \int dt \frac{G M_1 M_2}{r} \; .
    \end{aligned}\\[0.6cm]
    &\begin{aligned}
        \includegraphics[scale=0.8,valign=c]{BE_figure1.pdf}
        &=
        i\left(\frac{M_1}{2 \MPl}\right) \left(\frac{M_2}{2 \MPl}\right) \int dt_1 \int dt_2 \delta(t_1 - t_2) \int\frac{d^3\kbf}{(2\pi)^3} \frac{1}{\kabsbf^4} \\
        &\phantom{=-} \times \frac{\partial^2}{\partial t_1 \partial t_2} \exp(-i \kbf (\xbf_1(t_1) - \xbf_2(t_2))) P_{00;00} \\
        &= i  \frac{M_1 M_2}{8 \MPl^2} \vbf_{1,i} \vbf_{2,j} \int dt  \int\frac{d^3\kbf}{(2\pi)^3} \frac{\kbf_i \kbf_j}{\kabsbf^4} \exp(-i \kbf (\xbf_1(t_1) - \xbf_2(t_2)))  \\
        &= i \frac{M_1 M_2}{64 \pi \MPl^2} \int dt  \frac{1}{r} \left(\vbf_1 \cdot \vbf_2 - \frac{(\vbf_1 \cdot \rbf) (\vbf_2 \cdot \rbf)}{r^2}\right) =i \int dt \frac{G M_1 M_2}{2r} \left(\vbf_1 \cdot \vbf_2 - \frac{(\vbf_1 \cdot \rbf) (\vbf_2 \cdot \rbf)}{r^2}\right) \;.\\
    \end{aligned}\\[0.6cm]
    &\begin{aligned} 
        \includegraphics[scale=0.8,valign=c]{BE_figure2.pdf}
        &= i  \frac{M_1 M_2}{\MPl^2} \int dt_1\int dt_2 \delta(t_1 - t_2) \int\frac{d^3\kbf}{(2\pi)^3} \frac{\exp(-i \kbf (\xbf_1 - \xbf_2))}{\kabsbf^2} \\
        &\phantom{=-} \times \biggr[\underbrace{P_{0i;0j}}_{ = -\delta_{ij}/2} \vbf_{1i} \vbf_{2j} + \frac{1}{4} \biggr(\frac{\overbrace{P_{00;00}}^{ = 1/2}}{2} \vbf_1^2 + \underbrace{P_{ij;00}}_{ = \delta_{ij}/2} \vbf_{1,i} \vbf_{1,j}\biggr) \biggr]  \\
        &= -i \frac{M_1 M_2}{2\MPl^2} \int dt \int\frac{d^3\kbf}{(2\pi)^3} \frac{\exp(-i \kbf (\xbf_1 - \xbf_2))}{\kabsbf^2} \left(\vbf_{1} \cdot \vbf_{2} - \frac{3}{8} \vbf_1^2\right) \\
        &= -4i \int dt \frac{G M_1 M_2}{r} \left[\vbf_1 \cdot \vbf_2 - \frac{3}{8} \left(\vbf_1^2 + \vbf_2^2 \right) \right]\; .\\
    \end{aligned}\\[0.6cm]
    &\begin{aligned}
        \includegraphics[scale=0.8,valign=c]{BE_figure3.pdf}
        &= i \sumAB \left(\frac{M_1}{8 \MPl^2}\right) \left(\frac{M_2}{2 \MPl}\right) \left(\frac{M_2}{2 \MPl}\right) \int dt \int dt_1 \int dt_2 \delta(t-t_1) \delta(t-t_2) \\
        &\phantom{=-}\times \int\frac{d^3\kbf}{(2\pi)^3} \frac{\exp(-i \kbf (\xbf_1 - \xbf_2))}{\kabsbf^2} \int\frac{d^3\qbf}{(2\pi)^3} \frac{\exp(-i \qbf (\xbf_1 - \xbf_2))}{\qabsbf^2} P_{00;00}^2 \\
        &= \frac{i}{2} \sumAB \frac{M_1 M_2^2}{(32 \pi)^2 \MPl^4} \int dt \frac{1}{r}\\
        &= \frac{i}{2} \int dt \frac{G^2 M_1 M_2 (M_1 + M_2)}{r^2} \; .
    \end{aligned}\\[0.6cm]
    &\begin{aligned}
        \includegraphics[scale=0.8,valign=c]{BE_figure4.pdf}
        &= -\frac{i}{2} \sumAB \left(\frac{M_1}{2\MPl}\right) \left(\frac{M_2}{2\MPl}\right) \left(\frac{M_2}{2\MPl}\right)
            \left(\frac{1}{4\MPl}\right) \int dt \int dt_2 \int dt_3 \delta(t-t_2) \delta(t-t_3) \\
            &\phantom{=-} \times \int \frac{d^3\kbf_1}{(2\pi)^3} \int \frac{d^3\kbf_2}{(2\pi)^3} \int \frac{d^3\kbf_3}{(2\pi)^3} (2\pi)^3 \exp\left[i\kbf_1 x_1 + i(\kbf_2 + \kbf_3) x_2\right] \\
            &\phantom{=-} \times \delta^{(3)}\left(\kbf_1 + \kbf_2 + \kbf_3\right) \frac{\kbf_1^2 + \kbf_2^2 + \kbf_3^2}{|\kbf_1|^2 |\kbf_2|^2 |\kbf_3|^2} \\
            &= -\frac{i}{2} \sumAB \frac{M_1 M_2^2}{32 \MPl^4} \int dt \int \frac{d^3\kbf}{(2\pi)^3} \int \frac{d^3\qbf}{(2\pi)^3} \frac{|\kbf + \qbf|^2 + |\kbf|^2 + |\qbf|^2}{|\kbf + \qbf|^2  |\kbf|^2  |\qbf|^2} \exp\left(-i (\kbf+ \qbf) \rbf\right) \\
            &= -\frac{i}{2} \sumAB \frac{M_1 M_2^2}{32 \MPl^4} \Biggr[ \int dt \int \frac{d^3\kbf}{(2\pi)^3} \int \frac{d^3\qbf}{(2\pi)^3} \frac{1}{|\kbf|^2 |\qbf|^2} \exp\left(-i (\kbf+ \qbf) \rbf\right) \\
            &\phantom{=-} +\int dt \int \frac{d^3\kbf}{(2\pi)^3} \int \frac{d^3\qbf}{(2\pi)^3} \frac{|\kbf|^2 + |\qbf|^2}{|\kbf+\qbf|^2 |\kbf|^2 |\qbf|^2}  \exp\left(-i (\kbf+ \qbf) \rbf\right) \Biggr] \\
            &= -\frac{i}{2} \sumAB \frac{M_1 M_2^2}{32 \MPl^4}  \frac{1}{16 \pi^2} \int dt \frac{1}{r^2} = -i \int dt \frac{G^2 M_1 M_2 (M_1+M_2)}{r^2} \; ,\\[0.3cm] 
            &\text{where we drop the second term in the second to last line as it vanishes in dimensional} \\[-0.3cm] &\text{regularization.}
    \end{aligned}\\[0.6cm]
    &\begin{aligned} 
        \includegraphics[scale=0.8,valign=c]{BE_figure5.pdf}
        &= \frac{1}{2} \sumAB \left(\frac{- i q_1}{\MPl}\right)\left(\frac{- i q_2}{\MPl} \right)\int dt \int \frac{d^3\kbf}{(2\pi)^3} \frac{-i}{\kbf^2 + m_s^2} \exp\left(i \kbf \rbf\right) \left(1 - \vbf_1^2\right) \\
        &= \frac{1}{2} \sumAB i \frac{q_1 q_2}{4 \pi \MPl^2} \int dt \frac{\exp\left(-m_s r\right)}{r} \left(1 - \vbf_1^2\right)
        = i 8 G q_1 q_2 \int dt \frac{\exp\left(-m_s r\right)}{r} \left[1 - \frac{1}{2}\left(\vbf_1^2 + \vbf_2^2 \right)\right] \; ,
    \end{aligned}\\[0.6cm]
    &\begin{aligned}
        \includegraphics[scale=0.8,valign=c]{BE_figure6.pdf}
        &=
        \left(\frac{-i q_1}{\MPl}\right) \left(\frac{-i q_2}{\MPl}\right) \int dt_1 \int dt_2 \delta(t_1 - t_2) \int\frac{d^3\kbf}{(2\pi)^3} \frac{-i}{\left(\kabsbf^2 + m_s^2\right)^2} \\ 
        &\phantom{=-} \times \frac{\partial^2}{\partial t_1 \partial t_2} \exp(-i \kbf (\xbf_1(t_1) - \xbf_2(t_2)))  \\
        &= -i \frac{q_1 q_2}{\MPl^2} \vbf_{1,i} \vbf_{2,j} \int dt  \int\frac{d^3\kbf}{(2\pi)^3} \frac{\kbf_i \kbf_j}{\left(\kabsbf^2 + m_s^2\right)^2} \exp(-i \kbf (\xbf_1(t) - \xbf_2(t)))  \\
        &= -i \frac{q_1 q_2}{8 \pi \MPl^2} \int dt  \frac{\exp\left(-m_s r\right)}{r} \left(\frac{(\vbf_1 \cdot \rbf) (\vbf_2 \cdot \rbf)}{r^2} (1 + m_sr)-\vbf_1 \cdot \vbf_2\right) \\
        &= i 4 G q_1 q_2 \int dt  \frac{\exp\left(-m_s r\right)}{r} \left(\vbf_1 \cdot \vbf_2 - \frac{(\vbf_1 \cdot \rbf) (\vbf_2 \cdot \rbf)}{r^2} (1 + m_sr)\right)\;.\\
    \end{aligned}\\[0.6cm]
    &\begin{aligned}
        \includegraphics[scale=0.8,valign=c]{BE_figure7.pdf}
        &=  \sumAB \left( \frac{-i p_1}{\MPl^2} \right) \left( \frac{-i q_2}{\MPl}\right)\left( \frac{-i q_2}{\MPl}\right) \int dt \int dt_1 \delta(t - t_1) \int dt_2 \delta(t - t_2) \\
        &\phantom{=-}  \times \int\frac{d^3\kbf}{(2\pi)^3} \frac{-i \exp(-i \kbf (\xbf_1 - \xbf_2))}{\kabsbf^2+ m_s^2} \int\frac{d^3\qbf}{(2\pi)^3} \frac{-i \exp(-i \qbf (\xbf_1 - \xbf_2))}{\qabsbf^2+ m_s^2}  \\
        &= -i \sumAB \frac{p_1 q_2^2}{16 \pi^2 \MPl^4} \int dt  \frac{\exp(-2 m_s r)}{r^2} 
        = -i 64 G^2 \int dt (p_1 q_2^2 + p_2 q_1^2) \frac{\exp(-2 m_s r)}{r^2}
        \; .
    \end{aligned}\\[0.6cm]
    &\begin{aligned}
        \includegraphics[scale=0.8,valign=c]{BE_figure8.pdf}
        &= \sumAB \left(\frac{-i q_1}{2 \MPl^2}\right) \left( \frac{-i q_2}{\MPl} \right) \left( \frac{-i M_2}{2 \MPl} \right) \int dt \int dt_1 \delta(t-t_1)  \int dt_2 \delta(t-t_2) \\
        &\phantom{=-}  \times \int \frac{d^3\kbf}{(2\pi)^3} \frac{-i}{\kbf^2} \exp\left(i \kbf\rbf\right) P_{00;00} \int \frac{d^3\qbf}{(2\pi)^3} \frac{-i}{\qbf^2 + m_s^2} \exp\left(i \qbf\rbf\right) \\
        &=\sumAB -i \frac{q_1 q_2 M_2}{8 \MPl^4} \int dt \left(\frac{1}{4 \pi r}\right)^2 \exp(-m_s r) 
        = -i 8 G^2 q_1 q_2 (M_1+M_2) \int dt \frac{\exp(-m_s r)}{r^2}
        \; ,
    \end{aligned}\\[0.6cm]
    &\begin{aligned}
        \includegraphics[scale=0.8,valign=c]{BE_figure9.pdf}
        &= \sumAB \underbrace{(-i)^3 \left(\frac{- i q_2}{\MPl}\right)^2 \left( \frac{- i M_1}{2 \MPl} \right) \left( \frac{-i m_s^2}{4 \MPl} \right) \eta^{\mu \nu} P_{00;\mu\nu}}_{\equiv A} \int dt  \int \frac{d^3\kbf_1}{(2\pi)^3} \int \frac{d^3\kbf_2}{(2\pi)^3} \int \frac{d^3\kbf_3}{(2\pi)^3} \\
        &\phantom{=-}  \times \frac{e^{i \kbf_1 \rbf_1}}{\kbf_1^2 + m_s^2} \frac{e^{i \kbf_2 \rbf_1}}{\kbf_2^2 + m_s^2} \frac{e^{-i \kbf_3 \rbf_2}}{\kbf_3^2}  (2\pi)^3 \delta^{(3)}(\kbf_1 + \kbf_2 - \kbf_3)\\
        &= \sumAB A \int dt \int \frac{d^3\kbf_1}{(2\pi)^3} \int \frac{d^3\kbf_3}{(2\pi)^3} \frac{1}{\kbf_1^2 + m_s^2} \frac{1}{(\kbf_3 - \kbf_1)^2 + m_s^2} \frac{e^{i \kbf_3 \rbf}}{\kbf_3^2} \\[0.3cm]
        &\qquad \text{We employ the Feynman parametrization to rewrite the integral over } \kbf_1 \text{as}  \\[0.3cm]
        &\qquad \int \frac{d^3 \kbf_1}{(2\pi)^3} \frac{1}{\kbf_1^2 + m_s^2} \frac{1}{(\kbf_3 - \kbf_1)^2 + m_s^2} 
        = \int_0^1 da \int \frac{d^3 \kbf_1}{(2\pi)^3} \frac{1}{\left[ \kbf_1^2 + (a - a^2) \kbf_3^2 + m_s^2)  \right]^2}\\
        &\phantom{\qquad \int \frac{d^3 \kbf_1}{(2\pi)^3} \frac{1}{\kbf_1^2 + m_s^2} \frac{1}{(\kbf_3 - \kbf_1)^2 + m_s^2} }
        =\int_0^1 da \frac{1}{4 \pi \sqrt{(a-a^2) \kbf_3^2 + m_s^2}} \\
        &\phantom{\qquad \int \frac{d^3 \kbf_1}{(2\pi)^3} \frac{1}{\kbf_1^2 + m_s^2} \frac{1}{(\kbf_3 - \kbf_1)^2 + m_s^2} }= \frac{1}{2 \pi k_3} \arctan \left( \frac{k_3}{2m} \right) \; .\\
        &= \sumAB A \int dt \int \frac{d^3\kbf_3}{(2\pi)^3} \frac{1}{2 \pi k_3} \arctan \left( \frac{k_3}{2 m_s} \right) \frac{e^{i \kbf_3 \rbf}}{\kbf_3^2} \\[0.3cm]
        &\qquad  \text{This is solved by using spherical coordinates and partial integration.}\\[0.3cm]
        &= \sumAB (-i)^3 \left(\frac{- i q_2}{\MPl}\right)^2 \left( \frac{- i M_1}{2 \MPl}\right)\left( \frac{-i m_s^2}{4 \MPl}\right)  \underbrace{\eta^{\mu \nu} P_{00;\mu\nu}}_{=-1} \frac{ 1}{r} \frac{1}{(2\pi)^3} \\
        &\phantom{=-}  \times \int dt \left( - \frac{r \pi}{2} \text{Ei}(-2m_sr) + \frac{\pi}{4m_s} - \frac{\pi}{4m_s} e^{-2m_s r} \right)\\
        &= -i \frac{q_2^2 M_1 + q_1^2 M_2}{ 256 \pi^2 \MPl^4} \int dt \left( - 2m_s^2 \text{Ei}(-2m_sr) + \frac{m_s}{r} - \frac{m_s}{r} e^{-2m_sr} \right) \\ 
        &= -i 4 G^2 \left(q_2^2 M_1 + q_1^2 M_2 \right) \int dt \left( - 2m_s^2 \text{Ei}(-2m_sr) + \frac{m_s}{r} - \frac{m_s}{r} e^{-2m_sr} \right)
     \; .
    \end{aligned}\\[0.6cm]
    &\begin{aligned} 
        \includegraphics[scale=0.8,valign=c]{BE_figure10.pdf}
        &= \sumAB \underbrace{(-i)^3 \left( \frac{- i q_2}{\MPl} \right)\left(\frac{- i q_1}{\MPl}\right)\left(  \frac{- i M_2}{2 \MPl}\right)\left( \frac{-i m_s^2}{4 \MPl}\right) \eta^{\mu \nu} P_{00;\mu\nu}}_{A\equiv} \int dt \int \frac{d^3\kbf_1}{(2\pi)^3} \int \frac{d^3\kbf_2}{(2\pi)^3} \int \frac{d^3\kbf_3}{(2\pi)^3}\\
        &\phantom{=-}  \times \frac{e^{i \kbf_1 \rbf_1 }}{\kbf_1^2 + m_s^2} \frac{e^{i \kbf_2 \rbf_1}}{\kbf_2^2} \frac{e^{-i \kbf_3 \rbf_2}}{\kbf_3^2 + m_s^2} (2\pi)^3 \delta^{(3)}(\kbf_1 + \kbf_2 - \kbf_3)\\
        &= \sumAB A \int dt \int \frac{d^3\kbf_1}{(2\pi)^3} \int \frac{d^3\kbf_3}{(2\pi)^3} \frac{1}{\kbf_1^2 + m_s^2} \frac{1}{(\kbf_3 - \kbf_1)^2} \frac{e^{i \kbf_3 \rbf}}{\kbf_3^2 + m_s^2} \\
        &= \sumAB A \int dt \int \frac{d^3\kbf_3}{(2\pi)^3} \frac{1}{2 \pi k_3} \arctan \left( \frac{k_3}{m_s} \right) \frac{e^{i \kbf_3 \rbf}}{\kbf_3^2 + m_s^2}\\[0.3cm]
        &\text{At this point, we again switch to spherical coordinates and apply partial integration.} \\[0.3cm]
        &= \sumAB \frac{\pi}{r} \frac{A}{(2\pi)^3} \frac{1}{m_s} \int dt \underbrace{\frac{2}{\pi} \int_0^\infty dk' \arctan \left( k' \right) \frac{1}{k'^2 + 1} \sin (m_s r k')}_{\displaystyle\equiv \mathcal{I}(m_s r)} \; , \qquad \mathrm{with}  \quad  k' = \frac{|\kbf_3|}{m_s} \\
        &= i \frac{q_1 q_2 (M_1+M_2)}{64 \pi^2 \MPl^4} \frac{m_s}{r} \int dt \,\mathcal{I}(m_s r) \; \\
        &= -i 8 G^2 q_1 q_2 m_2 m_s \int dt \,\frac{1}{r} \bigg[ - \Ei(-2m_sr) e^{m_s r} + \log(2m_s r) e^{-m_s r} + \gamma_E e^{-m_s r} \bigg] \; .
    \end{aligned}\\[0.6cm]
    &\begin{aligned}
        \includegraphics[scale=0.8,valign=c]{BE_figure11.pdf}
        &= 
        \sumAB \overbrace{(-i)^3  \frac{3!}{2} \left( \frac{- i q_2}{\MPl} \right)^2 \left(\frac{- i q_1}{\MPl} \right) \left(\frac{-i c_3 \MPl }{3!}\right)}^{\equiv A} 
        \int dt \int d^3\rbf_v \int \frac{d^3\kbf_1}{(2\pi)^3} \int \frac{d^3\kbf_2}{(2\pi)^3} \int \frac{d^3\kbf_3}{(2\pi)^3} \\
         &\phantom{=-}  \times \frac{e^{i \kbf_1 (\rbf_1 - \rbf_v)}}{\kbf_1^2 + m_s^2} \frac{e^{i \kbf_2 (\rbf_1 - \rbf_v)}}{\kbf_2^2 + m_s^2} \frac{e^{i \kbf_3 (\rbf_v - \rbf_2)}}{\kbf_3^2 + m_s^2} 
         \\
         &= \sumAB (2 \pi)^3 A \int dt \int \frac{d^3\kbf_1}{(2\pi)^3} \int \frac{d^3\kbf_2}{(2\pi)^3} \int \frac{d^3\kbf_3}{(2\pi)^3} \frac{e^{i \kbf_1 \rbf_1}}{\kbf_1^2 + m_s^2} \frac{e^{i \kbf_2 \rbf_1}}{\kbf_2^2 + m_s^2} \frac{e^{-i \kbf_3 \rbf_2}}{\kbf_3^2 + m_s^2} \delta^{(3)}(\kbf_1+\kbf_2-\kbf_3) \\
         & = \sumAB A \int dt \int \frac{d^3\kbf_1}{(2\pi)^3} \int \frac{d^3\kbf_3}{(2\pi)^3} \frac{1}{\kbf_1^2 + m_s^2} \frac{1}{(\kbf_3 - \kbf_1)^2 + m_s^2} \frac{e^{i \kbf_3 \rbf}}{\kbf_3^2 + m_s^2}\\
         &= \sumAB A \int dt \int \frac{d^3\kbf_3}{(2\pi)^3} \frac{1}{2 \pi k_3} \arctan \left( \frac{k_3}{2m_s} \right) \frac{e^{i \kbf_3 \kbf}}{\kbf_3^2 + m_s^2} 
         \\
         &= \sumAB \frac{\pi}{r} \frac{A}{(2\pi)^3} \frac{1}{m_s} \int dt \underbrace{\frac{2}{\pi} \int_0^\infty dk' \arctan \left( k' \right) \frac{1}{4 k'^2 + 1} \sin (2 m_s r k')}_{\displaystyle \equiv \mathcal{K}(2 m_s r)} \\
         &= i \frac{q_2 q_1 (q_1 + q_2) c_3}{16 \pi^2 \MPl^2 m_s} \int dt \; \frac{\mathcal{K}(2 m_s r)}{r}  \\
        &= i \frac{q_2 q_1 (q_1 + q_2) c_3}{64 \pi^2 \MPl^2 m_s} \int dt \frac{1}{r} \left[-\Ei(-3 m_s r) e^{m_s r} + \Ei(-m_s r) e^{-m_s r} + \log(3) e^{-m_s r} \right] \\
        &=  i \frac{G}{2\pi} \frac{q_1 q_2 (q_1 + q_2) c_3}{m_s} \int dt \frac{1}{r} \left[-\Ei(-3 m_s r) e^{m_s r} + \Ei(-m_s r) e^{-m_s r} + \log(3) e^{-m_s r} \right] \; .
    \end{aligned}
\end{align}
\endgroup

\subsection{Scalar Radiation} \label{app:scalar_radiation_calculation}
Now we compute the diagrams that contribute to the radiative dynamics of the system. For the gravitational waveform at NLO, it is sufficient to consider the scalar radiation.

\begingroup
\addtolength{\jot}{.8em}
\begin{align}
    &\begin{aligned}
\includegraphics[scale=0.8,valign=c,raise=0.38cm]{R_figure5.pdf} = -i \sumAB \int dt \left( 1 - \frac{1}{2} \vbf_1^2 \right) \frac{q_1}{\MPl} \phiBar + \mathcal{O}(\vbf_1^4)
    \end{aligned}\\[0.6cm]
    &\begin{aligned}
\includegraphics[scale=0.8,valign=c,raise=0.38cm]{R_figure6.pdf}
        &= \sumAB -i \left( \frac{-i 2 p_1}{\MPl^2} \right) \left( \frac{-i q_2}{\MPl}\right) \int dt \int \frac{d^3 \kbf}{(2\pi)^3} \frac{e^{i \kbf (\mathbf{r_1} - \mathbf{r_2})}}{\kbf^2 + m_s^2} \phiBar = i \frac{p_1 q_2 + p_2 q_1}{2 \pi \MPl^2} \int dt \frac{e^{-m_s r}}{r} \frac{\phiBar}{\MPl}  
    \end{aligned}\\[0.6cm]
    &\begin{aligned}
\includegraphics[scale=0.8,valign=c,raise=0.38cm]{R_figure9.pdf}
    &= \sumAB (-i) \left( \frac{-i q_1}{2 \MPl^2} \right) \left( \frac{-i M_2}{2 \MPl} \right)  \int dt \int \frac{d^3 \kbf}{(2\pi)^3} \frac{e^{i \kbf (\mathbf{r_1} - \mathbf{r_2})}}{\kbf^2} \phiBar P_{00;00} \\
    & = i \frac{q_1 M_2 + q_2 M_1}{32 \pi \MPl^2} \int dt \frac{1}{r} \frac{\phiBar}{\MPl} \; . 
    \end{aligned}\\[0.6cm]
    &\begin{aligned} 
    \includegraphics[scale=0.8,valign=c]{R_figure7.pdf}
    &= \underbrace{3! \MPl (-i)^2 \left( \frac{-i q_1}{\MPl} \right) \left( \frac{-i q_2}{\MPl} \right) \left(\frac{-i c_3}{3!} \MPl \right)}_{A \equiv} \int dt \int d^3\rbf_v \int \frac{d^3\kbf_1}{(2\pi)^3} \frac{e^{i \kbf_1 (\rbf_1 - \rbf_v)}}{\kbf_1^2 + m_s^2} \\
    &\phantom{-=} \times \int \frac{d^3\kbf_2}{(2\pi)^3} \frac{e^{i \kbf_2 (\rbf_2 - \rbf_v)}}{\kbf_2^2 + m_s^2} \frac{\phiBar}{\MPl} \; .\\[0.3cm]
    &\text{Hence, the contribution of this diagram to the source term is} \\[0.3cm]
\mathmakebox[\widthof{\includegraphics[scale=0.8,valign=c]{R_figure7.pdf}}][r] J &= A \int \frac{d^3\kbf_1}{(2\pi)^3} \frac{e^{i \kbf_1 (\rbf_1 - \rbf_v)}}{\kbf_1^2 + m_s^2} \int \frac{d^3\kbf_2}{(2\pi)^3} \frac{e^{i \kbf_2 (\rbf_2 - \rbf_v)}}{\kbf_2^2 + m_s^2} \; .\\[0.3cm]
        &\text{With this we can use Eq.~\eqref{eq::multipole_moments} to calculate the $l=1$ and $p=0$ moment.} \\[0.3cm]
\mathmakebox[\widthof{\includegraphics[scale=0.8,valign=c]{R_figure7.pdf}}][r]{\mathcal{I}^i}
        &= A \int d^3\rbf_v \int\frac{d^3\kbf_1}{(2\pi)^3} \int\frac{d^3\kbf_2}{(2\pi)^3} \frac{e^{i \kbf_1(\rbf_1 - \rbf_v)}}{\kbf_1^2 + m_s^2} \frac{e^{\kbf_2(\rbf_2 - \rbf_v)}}{\kbf_2^2 + m_s^2} \rbf_v^i \\
        &= A  \int d^3\rbf_v  \int\frac{d^3\kbf_1}{(2\pi)^3} \int\frac{d^3\kbf_2}{(2\pi)^3} \frac{e^{i \kbf_1(\rbf_1 - \rbf_v)}}{\kbf_1^2 + m_s^2} \left(i \frac{2 \kbf_2^i}{(\kbf_2^2 + m_s^2)^2} e^{i \kbf_2 (\rbf_2 - \rbf_v)} + \frac{\rbf_2^i}{\kbf_2^2 + m_s^2} e^{i \kbf_2 (\rbf_2  - \rbf_v)}\right) \\
        &= A  \int \frac{d^3\kbf_1}{(2\pi)^3} \frac{e^{i\kbf_1\rbf}}{\kbf_1^2 + m_s^2} \left(-i \frac{2 \kbf_1^i}{(\kbf_1^2 + m_s^2)^2} + \frac{\rbf_2^i}{\kbf_1^2 + m_s^2}\right) \\
        &= A  \frac{1}{16 m_s \pi} e^{- m_s r} (\rbf_1^i + \rbf_2^i) \\
        &= -i \frac{q_1 q_2 c_3}{16 m_s \pi} e^{- m_s r} (\rbf_1^i + \rbf_2^i) \; .
    \end{aligned}\\[0.6cm]
        &\begin{aligned}
    \includegraphics[scale=0.8,valign=c]{R_figure8.pdf}
        &= \sumAB \underbrace{(-i)^2 \MPl \left(\frac{-i q_1}{\MPl} \right) \left( \frac{-i M_2}{2 \MPl} \right) \left( \frac{-i 2 m_s^2}{4 \MPl}\right)}_{\equiv A} \int dt \int d^3 \rbf_v \int \frac{d^3 \kbf_1}{(2\pi)^3} \frac{e^{i \kbf_1 (\rbf_1 - \rbf_v)}}{\kbf_1^2 + m_s^2} \\
        &\phantom{=-}  \times \int \frac{d^3 \kbf_2}{(2\pi)^3} \frac{e^{i \kbf_2 (\rbf_v - \rbf_2)}}{\kbf_2^2} \eta^{\mu\nu} P_{\mu\nu; 00} \frac{\phiBar}{\MPl} \;.\\[0.3cm]
        \qquad &\text{Again we identify the contribution of this diagram to the source term:} \\[0.3cm]
\mathmakebox[\widthof{\includegraphics[scale=0.8,valign=c]{R_figure8.pdf}}][r] J &= -A \int \frac{d^3\kbf_1}{(2\pi)^3} \frac{e^{i \kbf_1 (\rbf_1 - \rbf_v)}}{\kbf_1^2 + m_s^2} \int \frac{d^3\kbf_2}{(2\pi)^3} \frac{e^{i \kbf_2 (\rbf_2 - \rbf_v)}}{\kbf_2^2} \; .\\[0.3cm]
        \qquad &\text{From this we obtain the $l=1$ and $p=0$ moment via Eq.~\eqref{eq::multipole_moments}.}\\[0.3cm]
\mathmakebox[\widthof{\includegraphics[scale=0.8,valign=c]{R_figure8.pdf}}][r]{\mathcal{I}^i} 
        &= \sumAB -A \int d^3\rbf_v \int\frac{d^3\kbf_1}{(2\pi)^3} \int\frac{d^3\kbf_2}{(2\pi)^3} \frac{e^{i \kbf_1(\rbf_1 - \rbf_v)}}{\kbf_1^2 + m_s^2} \frac{e^{\kbf_2(\rbf_2 - \rbf_v)}}{\kbf_2^2} \rbf_v^i \\
        &= \sumAB -A  \int d^3\rbf_v  \int\frac{d^3\kbf_1}{(2\pi)^3} \int\frac{d^3\kbf_2}{(2\pi)^3} \frac{e^{i \kbf_1(\rbf_1 - \rbf_v)}}{\kbf_1^2 + m_s^2} \left(i \frac{2 \kbf_2^i}{|\kbf_2|^4} e^{i \kbf_2 (\rbf_2 - \rbf_v)} + \frac{\rbf_2^i}{\kbf_2^2} e^{i \kbf_2 (\rbf_2  - \rbf_v)}\right) \\
        &= \sumAB -A  \int \frac{d^3\kbf_1}{(2\pi)^3} \frac{e^{i\kbf_1\rbf}}{\kbf_1^2 + m_s^2} \left(-i \frac{2 \kbf_1^i}{|\kbf_1|^4} + \frac{\rbf_2^i}{\kbf_1^2}\right) \\
        &= \sumAB -A  \left[\rbf^i  \frac{(m_s^2 r^2 - 2) + 2e^{-m_s r}(m_s r+1)}{4 \pi m_s^4 r^3} + \rbf_2^i \frac{1 - e^{-m_s r}}{4\pi m_s^2 r}\right] \\
        &= \sumAB i \frac{q_1 M_2}{4 \MPl^2}  \left[\rbf^i  \frac{(m_s^2 r^2 - 2) + 2e^{-m_s r}(m_s r+1)}{4 \pi m_s^2 r^3} + \rbf_2^i \frac{1 - e^{-m_s r}}{4\pi r}\right] \; .
    \end{aligned}
\end{align}
\endgroup

\cftaddtitleline{toc}{section}{\textbf{References}}{\thepage}

\let\oldaddcontentsline\addcontentsline
\renewcommand{\addcontentsline}[3]{}
\bibliography{biblio.bib}
\let\addcontentsline\oldaddcontentsline
\end{document}